\documentclass[epsfig,12pt]{article}
\usepackage{amssymb}
\usepackage{epsfig}

\begin{document}

\title{\vspace{-0.9in}\large \bf Convergence of Fine-lattice Discretization for Near-critical Fluids}
\vspace{-0.25in}

\author{\normalsize Sarvin Moghaddam,$^{1,2}$ Young C. Kim,$^{2}$ and Michael E. Fisher$^{2}$ \\
$^{1}$\normalsize Department of Chemical Engineering, Princeton University, Priceton, New Jersey 08544,\\
$^{2}$\normalsize Institute for Physical Science and Technology, University of Maryland,\\ \normalsize College Park, Maryland 20742}

\date{\normalsize\today}

\maketitle
\vspace{-0.2in}
\begin{abstract}

In simulating continuum model fluids that undergo phase separation and criticality, significant gains in computational efficiency may be had by confining the particles to the sites of a lattice of sufficiently fine spacing, $a_{0}$ (relative to the particle size, say $a$). But a cardinal question, investigated here, then arises, namely: How does the choice of the lattice discretization parameter, $\zeta\equiv a/a_{0}$, affect the values of interesting parameters, specifically, critical temperature and density, $T_{\mbox{\scriptsize c}}$ and $\rho_{\mbox{\scriptsize c}}$? Indeed, for small $\zeta\, (\lesssim 4\,$-$\,8)$ the underlying lattice can strongly influence the thermodynamic properties. A heuristic argument, essentially exact in $d=1$ and $d=2$ dimensions, indicates that for models with hard-core potentials, both $T_{\mbox{\scriptsize c}}(\zeta)$ and $\rho_{\mbox{\scriptsize c}}(\zeta)$ should converge to their continuum limits as $1/\zeta^{(d+1)/2}$ for $d\leq 3$ when $\zeta\rightarrow\infty$; but the behavior of the error is highly erratic for $d\geq 2$. For smoother interaction potentials, the convergence is faster. Exact results for $d=1$ models of van der Waals character confirm this; however, an optimal choice of $\zeta$ can improve the rate of convergence by a factor $1/\zeta$. For $d\geq 2$ models, the convergence of the {\em second virial coefficients} to their continuum limits likewise exhibit erratic behavior which is seen to transfer similarly to $T_{\mbox{\scriptsize c}}$ and $\rho_{\mbox{\scriptsize c}}$; but this can be used in various ways to enhance convergence and improve extrapolation to $\zeta = \infty$ as is illustrated using data for the restricted primitive model electrolyte.

\end{abstract}
\pagebreak

\section{Introduction and General Considerations}
\indent

Understanding the thermodynamic behavior of complex fluids, such as electrolytes, polymers, colloids, etc., has been a key issue in soft-condensed-matter physics in recent years. Simulating realistic continuum models of such fluids with the aid of powerful computers can be informative; however, very large computing resources may be required. Even determining accurately the critical parameters of the simplest model of an electrolyte, namely the restricted primitive model (RPM), has been an outstanding problem for more than a decade.$^{1}$ To ease the difficulties of simulation, various computer algorithms and special techniques have been developed. Particularly helpful is the fine-lattice discretization approach introduced by Panagiotopoulos and Kumar$^{2}$ that has provided an efficient way of simulating the thermodynamic properties near phase separation and criticality in various model fluids.$^{2-8}$ In this method, as illustrated in Figure 1, the center of any particle may reside only on the sites of a lattice of sufficiently fine spacing $a_{0}$, relative to the particle size, $a$, which it is usually convenient to take as the diameter of the intermolecular repulsive core. A crucial parameter is then the lattice-discretization ratio which we define generally by
 \begin{equation}
  \zeta\equiv a/a_{0} \geq 1,
 \end{equation} 
where, as is worth stressing at the outset, $\zeta$ need not be an integer. Systems with $\zeta=1$ correspond to the simplest lattice gas models with, typically, single-site hard cores that merely prevent double occupancy of a lattice site. On the other hand, when $\zeta\rightarrow\infty$ one clearly recaptures the full continuum version of the model fluid under study. In practice one observes that even for moderately low values of $\zeta$, say in the range 5$\,$-10, the properties of the discretized systems are close to those of their continuum limits.$^{3,9}$

The advantage of simulating such discretized systems rather than their continuum counterparts is that one can readily generate a look-up table of the interactions between two particles at all allowable distances (up to a maximum of order the system size, $L$) prior to undertaking the simulations. This then speeds up the computations markedly: thus for the discretized RPM one gains a factor of speed as large as 100 relative to the continuum model.$^{3}$ On the other hand, one drawback of the approach is that one cannot simulate using arbitrarily large values of $\zeta$ owing to the limitations on the capacity of fast-access memory in which to store the interaction data. Hence, some compromise is inescapable --- especially when large system-sizes, $L$, are demanded$^{10}$ --- in order to select a reasonably low value of $\zeta$ that will still ensure satisfactory estimates of continuum properties.

To this end we hence ask: How does the lattice-discretization parameter, $\zeta$, affect the thermodynamics of the model fluids under study? Of particular interest, and a focus of this article, is the $\zeta$-dependence of the critical parameters of systems that undergo phase separation and gas-liquid or fluid-fluid criticality. In particular, how do the critical temperature and density, $T_{\mbox{\scriptsize c}}$ and $\rho_{\mbox{\scriptsize c}}$, depend on $\zeta$? More specifically, one may expect that $T_{\mbox{\scriptsize c}}(\zeta)$ and $\rho_{\mbox{\scriptsize c}}(\zeta)$ will converge to their continuum limits as $1/\zeta^{\psi_{T}}$ and $1/\zeta^{\psi_{\rho}}$, respectively, with convergence exponents $\psi_{T}$ and $\psi_{\rho}$ that are clearly of interest. It is, indeed, reasonable to expect $\psi_{T}=\psi_{\rho}=\psi$ for general cases; but, how does $\psi$ depend on the details of the models? Furthermore, understanding these issues may provide a route to reliably estimating properties such as $T_{\mbox{\scriptsize c}}$ and $\rho_{\mbox{\scriptsize c}}$ precisely via an extrapolation to the continuum limit $(\zeta\rightarrow\infty)$ based on simulations at moderately low values of $\zeta$.

Numerical studies of the $\zeta$-dependence of $T_{\mbox{\scriptsize c}}$ and $\rho_{\mbox{\scriptsize c}}$ have previously been performed for certain model fluids.$^{4,5,8}$ For the RPM --- hard-core ions carrying charges $\pm q_{0}$ --- Panagiotopoulos$^{4}$ concluded that both the critical temperature and density approach their continuum limits approximately as $1/\zeta^{2}$ for values of $\zeta$ from 5 up to 15. On the other hand, uncharged models with strong but smoothly rising, $e^{-cr}$ repulsive interactions have exhibited faster rates of convergence of $T_{\mbox{\scriptsize c}}(\zeta)$; namely, the deviations $T_{\mbox{\scriptsize c}}(\zeta)-T_{\mbox{\scriptsize c}}(\infty)$ appeared to scale as $1/\zeta^{\psi}$ with $\psi=6\pm 2$ for $\zeta$ in the range $(5,\,10)$.$^{5}$ [Moreover, for the models examined the estimates of $\rho_{\mbox{\scriptsize c}}(\zeta)$ displayed rather small dependence on $\zeta$.] In other calculations,$^{5}$ it was observed that a charged dumbbell model with hard-core repulsions exhibited the same convergence rate as the RPM, namely with an exponent $\psi\simeq 2$. Later, rather more precise simulations$^{8}$ confirmed the approximate overall $1/\zeta^{2}$ convergence of $T_{\mbox{\scriptsize c}}$ and $\rho_{\mbox{\scriptsize c}}$ for the RPM but the detailed variation with $\zeta$ was, in fact, found to be quite erratic or `noisy' at a level significantly larger than the simulation uncertainties. How can one understand these numerical observations from an analytical or general theoretical viewpoint?

To answer this question, it is essential to understand the nature of the truncation errors of the thermodynamic and statistical properties of the discretized system relative to its continuum counterpart as $\zeta$ changes. More concretely, in the grand canonical ensemble, for example, the partition function, $\Xi (T,\mu)$, can be obtained by integrating the Boltzmann factor for the $N$-particle Hamiltonian ${\cal H}_{N}$ over the full phase space and by summing on $N$. On the other hand, in the discretized system, the configuration space is limited to the lattice points of spacing $a_{0}=a/\zeta$; thus the spatial integrations defining $\Xi$ must be replaced by appropriate summations over lattice configurations, yielding a $\Xi^{\zeta} (T,\mu)$ which deviates from the continuum limit, $\Xi^{\infty} (T,\mu)$. The grand thermodynamic potential from which the thermodynamic quantities can be derived is then proportional to $\ln\Xi$. How does $\zeta$ affect the partition function (and thus the corresponding thermodynamics) for the discretized system?

To simplify the situation and to gain a first insight into the character of the $\zeta$-convergence of $T_{\mbox{\scriptsize c}}$ and $\rho_{\mbox{\scriptsize c}}$, it is instructive, following ref 8, to examine an approximate, van der Waals (vdW) description of the equation of state for a model fluid. Accordingly, for a $d$-dimensional fluid with pair interaction potential $\varphi(r)=\varphi_{0}(r)+\varphi_{1}(r)$, where $\varphi_{0}$ is repulsive and of short range, while $\varphi_{1}$ is smooth, attractive and long ranged, let us consider the vdW-type equation of state,
 \begin{equation}
  p/\rho k_{B}T = [1-\rho B_{0}(T)]^{-1} + \rho B_{1}(T),   \label{eq1}
 \end{equation}
which will be valid semiquantitatively and, in general, exhibit (classical) criticality. In this equation $B_{0}(T)$ and $B_{1}(T)$ are partial second virial coefficients given by
 \begin{equation}
   B_{i}(T) = -\mbox{$\frac{1}{2}$}\int d^{d}r f_{i}(r), \hspace{0.2in} 1+f_{i} = e^{-\varphi_{i}(r)/k_{B}T}, \hspace{0.2in} \mbox{for $i=0,1$}, \label{eq.3}
 \end{equation}
while for the full second virial coefficient, $B_{2}(T)=B_{0}(T)+B_{1}(T)$, one simply replaces $\varphi_{i}(r)$ by the total potential $\varphi(r)$. For a discretized system with $\zeta=a/a_{0}$, the $B_{i}$ integrals must be replaced by sums, 
 \begin{equation}
  B_{i}^{\zeta} = -\mbox{$\frac{1}{2}$}\sum_{\mbox{\boldmath\scriptsize $n$}}f_{i}(\mbox{\boldmath $n$}a_{0})a_{0}^{d}, \label{eq.4}
 \end{equation}
over integral lattice vectors, {\boldmath $n$}, and likewise for $B_{2}^{\zeta}(T)$. The critical temperature and density, $T_{\mbox{\scriptsize c}}(\zeta)$ and $\rho_{\mbox{\scriptsize c}}(\zeta)$, are then readily derived from $B_{0}^{\zeta}(T)$ and $B_{1}^{\zeta}(T)$. Thus it is straightforward to see that the convergence of the $B_{i}^{\zeta}(T)$ to their continuum limits, $B_{i}^{\infty}(T)$ --- characterized, say, by an exponent $\psi_{B}$ --- rather directly governs the behavior of $T_{\mbox{\scriptsize c}}(\zeta)$ and $\rho_{\mbox{\scriptsize c}}(\zeta)$. In particular, in the absence of special or ``accidental'' cancellations, one will have $\psi_T = \psi_\rho = \psi_B$.

Of course, we do not expect this approach to yield quantitatively accurate information since the true critical behavior in two- and three-dimensional systems deviates from the classical, vdW paradigm. However, we do expect the results to be correct at a semiquantitative level as, indeed, are the estimates of $T_{\mbox{\scriptsize c}}$ and $\rho_{\mbox{\scriptsize c}}$ obtained from the van der Waals equation itself.$^{1}$ Similarly, we anticipate, and will demonstrate numerically and analytically, that the character of the asymptotic behavior as embodied in the exponents $\psi$, etc., is in general preserved.

Nevertheless, it is not unreasonable to question whether the existence of the nonclassical critical exponents $\beta<\frac{1}{2}$, $\gamma >1$, $\nu > \frac{1}{2}$, etc., for the true thermodynamic coexistence curve, compressibility, and correlation length, etc., might not change the convergence exponents $\psi$ for $T_{\mbox{\scriptsize c}}(\zeta)$ and $\rho_{\mbox{\scriptsize c}}(\zeta)$. This is an interesting question which we address in Appendix A. The arguments there indicate that in leading order when $\zeta$ increases, the behavior of the critical parameters will be dominated by the convergence of low-order integrals over Boltzmann factors, as epitomized in the virial coefficients, and thus will in be insensitive generically to the nonclassical subtleties of the true critical behavior.$^{11}$

To follow the line of investigation proposed we must thus ask: How rapidly will the $B_{i}^{\zeta}(T)$ converge to their limits? Specifically, if we define the relative truncation error$^{8}$ via
 \begin{equation}
  E_{i}(\zeta;T) = B_{i}^{\zeta}(T)/B_{i}^{\infty}(T) - 1,  \label{eq.5}
 \end{equation}
and consider the approximation of the integrals (\ref{eq.3}) by the sums (\ref{eq.4}), it is rather clear that the rate of decay of $E_{i}(\zeta)$ to 0 as $\zeta$ increases will be determined by the continuity and smoothness of the integrands $f_{i}(r)$. In particular, if $f_{i}(r)$ or any of its low-order derivatives exhibit discontinuities, we should expect these to play a dominant role. It follows at once from our description of $\varphi_{0}(r)$ and $\varphi_{1}(r)$ that the strongly varying repulsive potentials and, thence, $E_{0}(\zeta)$, should be the major players. Indeed, some of the explicit calculations we present below will amply illustrate this inference. Already we can see why the RPM and the hard-core dumbbell models should, by virtue of the corresponding discontinuities in the relevant Mayer $f(r)$ functions, have been expected to exhibit a slower convergence than the uncharged models with smooth repulsive potentials.

Because they lead to the slowest convergence and because of their ubiquity in theoretical modeling (albeit primarily for analytical and conceptual convenience) let us focus on the effects of hard-core repulsive potentials. A few moments thought shows that $2B_{0}^{\zeta}(T)/a_{0}^{d}$ is just the number of lattice points, say $N_{d}(\zeta)$, contained within a $d$-dimensional sphere centered on a lattice site and of radius $\zeta$ (using units of the lattice spacing $a_{0}$): see Figure 1. Likewise $2B_{0}^{\infty}(T)/a_{0}^{d}$ is the volume of the corresponding sphere, say $V_{d}(\zeta)$. If, as for simplicity we will suppose henceforth, attention is confined to square, simple cubic, and hypercubic lattices [with lattice vectors $\mbox{\boldmath $n$}=(n_{r})$ $(r=1,2,\cdots,d)$ and $n_{r}=0,\pm 1,\pm 2,\cdots$], one soon realizes that the behavior of $N_{d}(\zeta)$ as $\zeta$ increases is a classic problem in the geometry of numbers.$^{12}$

A heuristic argument, presented in Appendix B for the insight we feel it conveys, suggests that the truncation error $E_{0}(\zeta)$ decays as $1/\zeta^{(d+1)/2}$ for $d\leq 3$, while for $d > 3$ its decay follows a $1/\zeta^{2}$ power law, independent of $d$. This argument can be checked immediately for $d=1$ since for any non-integral $\zeta$ in a linear lattice we have $|V_{1}(\zeta)-N_{1}(\zeta)|\leq 2$ while $V_{1}(\zeta)= 2\zeta$ so that $E_{0}(\zeta)$ decays as $1/\zeta$. On the other hand, for $d\geq 4$ it has been proven rigorously$^{13}$ that $E_{0}(\zeta)$ decays as $1/\zeta^{2}$ whatever the value of $d$.$^{12,14}$ For $d=2$, Bleher, with Dyson and Lebowitz,$^{15,16}$ has established that $E_{0}(\zeta)$ is of magnitude $1/\zeta^{3/2}$ but has an erratic or noisy behavior. Finally, Bleher and Dyson$^{14,16}$ have shown that $d=3$ is a borderline dimension at which $E_{0}(\zeta)$ contains a logarithmic factor and, indeed, varies in magnitude as $(\ln\zeta)^{1/2}/\zeta^{2}$. These results confirm the heuristic arguments --- see section II and Appendix B --- except for the appearance of the factor $\sqrt{\ln\zeta}$ when $d=3$: this presents an interesting analogy with the theory of critical phenomena where the free energy and other properties invariably contain logarithmic factors at borderline dimensions.

Indeed, as we have indicated, the convergence of $T_{\mbox{\scriptsize c}}(\zeta)$ and $\rho_{\mbox{\scriptsize c}}(\zeta)$ to their limits observed numerically in the RPM and the hard-core dumbbell model agrees with our analyses. However, the logarithmic correction factor is too difficult to detect, in light of the limited (small integer) values of $\zeta$ examined. The faster convergence of $T_{\mbox{\scriptsize c}}(\zeta)$ and $\rho_{\mbox{\scriptsize c}}(\zeta)$ in the uncharged models$^{5}$ may clearly be attributed to the smoothness of the repulsive potentials adopted. On general grounds, well illustrated in section III for $(d\,$$=$$\,1)$-dimensional systems, one expects that as the integrands $f_{0}(r)$ and $f_{1}(r)$ exhibit discontinuities in successively higher derivatives $0,1,2,\cdots\,$, so the truncation errors will decrease by further factors of $1/\zeta$. Thus for Boltzmann factors $\,\,\exp [-\varphi(r)/k_{\mbox{\scriptsize B}}T]\,\,$ with continuous second derivatives one might expect truncation errors decreasing faster than $1/\zeta^{4}$ for $d=3$ dimensions. This argument must be viewed with caution, however, even as regards potentials with many or all derivatives continuous: indeed, it is most probable that the observed exponent estimates for these smooth-potential models, namely $\psi= 6\pm 2$, represent {\em effective} (rather than truly asymptotic) {\em exponents} that are relevant to the finite range of $\zeta$ explored and to the relative sharpness of the Boltzmann-factor variation: see section IV.

The balance of this article is organized as follows. In section II the behavior of the partial second virial coefficient $B_{0}^{\zeta}$ for a hard-core potential is discussed in further detail focusing, in particular, on $d=2$ and $3$. In section III, following Fisher and Widom,$^{17}$ exact results for the critical parameters of one-dimensional discretized van der Waals models are obtained for various short-range potentials. The convergence of the calculated values of $T_{\mbox{\scriptsize c}}(\zeta)$ and $\rho_{\mbox{\scriptsize c}}(\zeta)$ demonstrates in explicit fashion the role of discontinuities and the smoothness of the pair potentials. The truncation errors in two and three dimensions are considered in section IV with illustrations based, in particular, on simulation data for the RPM.$^{8}$ A brief summary is presented in section V.

\section{Hard-core Second Virial Coefficient}
\indent

For subsequent applications to extrapolating to the continuum limit the critical points of the discretized RPM, or other model fluids with hard-core interactions, we consider in this section the hard-core part of the second virial coefficient, namely, $B_{0}^{\zeta}$. As pointed out, this is equivalent, via suitable normalization, to determining the number of lattice points $N_{d}(\zeta)$ inside a $d$-dimensional sphere of radius $R=\zeta a_{0}$. When $\zeta\rightarrow\infty$, it is obvious that $N_{d}(\zeta)$ increases like the reduced volume, $V_{d}(\zeta)\sim \zeta^{d}$, of the sphere; but the crucial question is: How does $N_{d}(\zeta)$ approach the continuum behavior? One may write
 \begin{equation}
   N_{d}(\zeta) = V_{d}(\zeta) + n_{d}(\zeta),
 \end{equation}
where the deviation $n_{d}(\zeta)$ satisfies $n_{d}(\zeta)/\zeta^{d}\rightarrow 0$ as $\zeta\rightarrow\infty$. More specifically we wish to know how $n_{d}(\zeta)$ depends on $\zeta$. 

For $d$$\,=\,$$1$, it is trivial that $n_{d}$$\,\sim\,$$ O(1)$ so that $E_{0}(\zeta)$$\,=\,$$n_{d}(\zeta)/V_{d}(\zeta)$ defined as in eq \ref{eq.5} decays as $1/\zeta$: see Figure 2(a) which displays a scaled plot. Note, first, that $E_{0}(\zeta)$ is a piecewise continuous function, locally always decreasing but with infinitely many positive jump discontinuities. These features, in fact, characterize the variation of $E_{0}(\zeta)$ for all $d$: see Figure 2(b). Second, observe that $E_{0}(\zeta)$ vanishes when $\zeta$ takes half-integer values: this, of course, is special to $d$$\,=\,$$1$. In the next section these values for $\zeta$ will be shown to provide faster convergence for the critical parameters of discretized one-dimensional models. A selection of similar special $\zeta$ values for $d$$\,=\,$$2$ and $d$$\,=\,$$3$ dimensions is presented in Table I below.

To understand the convergence of the truncation error $E_{0}(\zeta)$ for $d>1$, consider a $d$-dimensional ball of radius $\zeta$ centered at the origin site of a simple hypercubic lattice. The error term, $n_{d}(\zeta)$, then arises near the surface of the ball which cuts elementary lattice cells into two parts. A heuristic argument presented in Appendix B supposes that the cutting of nearby cells is only weakly correlated: in that case the root mean square magnitude of $n_{d}(\zeta)$ grows like $\zeta^{(d-1)/2}$. This implies that $E_{0}(\zeta)$ should decay as $1/\zeta^{(d+1)/2}$. This argument, however, is inadequate for $d>3$ where the exact results show that $E_{0}(\zeta)$ decays only as $1/\zeta^{2}$ with a convergence exponent independent of $d$ as recounted in section I.$^{12}$ In other words, the error function $n_{d}(\zeta)$ grows as $\zeta^{d-2}$ for $d>3$, i.e., much faster than $\zeta^{(d-1)/2}$ as suggested by the heuristic argument with limited correlations between nearby cells. Indeed, as demonstrated in the second part of Appendix B, the reason for the distinct behavior is that the surface of a sphere is effectively flatter in higher dimensions $(d>3)$ so that the lattice cells cut by the sphere are strongly correlated over regions of the sphere having $d$$\,-\,$$2$ dimensions of order the radius $R$.

For our present purposes the most interesting cases are $d$$\,=\,$$2$ and $3$, as appropriate for realistic model fluids. For $d$$\,=\,$$2$, Bleher, Dyson and Lebowitz$^{15}$ demonstrated some time ago, in a completely different physical context, that $n_{2}(\zeta)$ fluctuates rather wildly growing in magnitude as $\zeta^{1/2}$. The truncation error, $E_{0}(\zeta)$, then decays as $1/\zeta^{3/2}$ as illustrated in Figure 2(b). This is, indeed, in accord with our heuristic argument. However, Bleher {\em et al.}$^{15}$ went further by showing that the scaled fluctuation, $n_{2}(\zeta)/\zeta^{1/2}$, takes on values in the interval between $s$ and $s+ds$ with a probability density, ${\cal P}(s)$, that converges to a non-Gaussian limiting distribution, say ${\cal P}_{\infty}(s)$, that decays  roughly as $\exp(-c s^{4})$ for large $s$.$^{15}$

For the three-dimensional $(d=3)$ case, Bleher and Dyson$^{13,14}$ established that $n_{3}(\zeta)$ grows as $\zeta\sqrt{\ln\zeta}$ when $\zeta\rightarrow\infty$. The behavior of $s=n_{3}(\zeta)/\zeta\sqrt{\ln\zeta}$ as a function of $\zeta$, exhibits, just as for $d=2$, essentially random fluctuations about a zero mean: For a depiction of the behavior of the wild variation of the truncation error, $E_{0}(\zeta)$, for $d=3$ see Figure 3 of ref 8. Bleher and Dyson also conjectured on the basis of a heuristic argument that the distribution of the variable, $s=n_{3}(\zeta)/\zeta\sqrt{\ln\zeta}$, would approach a Gaussian distribution. To illustrate this conjecture numerically, we present in Figure 3 the finite-$\zeta$ distributions, ${\cal P}_{\zeta}(s)$, obtained for $\zeta=5$ up to $1000$ using increments $\Delta\zeta = 10^{-4}$. For $\zeta\geq 100$, the distribution is already very close to the limiting distribution which is also shown using an estimated variance $\sigma^{2}=6.85$. To check that the distribution is indeed a Gaussian we have computed the ratio $\langle s^{4}\rangle/\langle s^{2}\rangle^{4}$ obtaining the values $2.52$, $2.77$, $2.89$, and $3.00$ for the $\zeta$ ranges shown in Figure 3. For a Gaussian distribution, this ratio should be exactly 3.

The appearance of a logarithmic factor in $s$ (not predicted by our heuristic arguments) would normally be associated with an extra parameter, i.e., $\ln\zeta\Rightarrow\ln c_{0}\zeta = (\ln\zeta +\ln c_{0})$. It is not so surprising, therefore, that the data of Fig.\ 3 even for $\zeta$$\,\geq\,$$100$ show systematic deviations from symmetry around $s$$\,=\,$$0$ that are consistent with a value of $c_{0}$$\,>\,$$1$.

Finally, we may recall that for $d$$\,>\,$$3$ the truncation errors always decrease in magnitude as $1/\zeta^{2}$; and, in analogy with critical phenomena, one might speculate that the errors will always be distributed in Gaussian fashion.

\section{Exact Results for One-dimensional Models}

\subsection{Motivation}
\indent

Here we consider exactly solvable $(d=1)$-dimensional fluid models in a discretized space and calculate the critical parameters as functions of $\zeta=a/a_{0}$. Recall again that $a$ measures the size of particles which, in this section will always be the hard-core diameter, while $a_{0}$ is the lattice spacing of the discretized system. We will verify explicitly that discontinuities in $f_{i}(r)$ or its derivatives dominate the convergence of the critical parameters $T_{\mbox{\scriptsize c}}(\zeta)$ and $\rho_{\mbox{\scriptsize c}}(\zeta)$ when $\zeta\rightarrow\infty$.

Now, if $~\Xi(\beta,z;L)$ is the grand canonical partition function of a linear one-dimensional system of length $L$ at temperature $T$$\,=\,$$1/k_{\mbox{\scriptsize B}}\beta$ and activity $z$, the appropriate one-dimensional ``pressure'' in the thermodynamic limit is given by
  \begin{equation}
   \beta p = \lim_{L\rightarrow\infty}(1/L)\ln \Xi(\beta,z;L).  \label{eq4.1}
  \end{equation}
This limit can be most efficaciously calculated via the construction of the Laplace transform
  \begin{equation}
   \Psi(\beta,z;s) = \int_{0}^{\infty} e^{-sL}\Xi(\beta,z;L)dL, \label{eq4.2}
  \end{equation}
where the integral is absolutely convergent for all values of $s$ with a real part exceeding the abscissa of convergence, $s_{0}(\beta,z)$.$^{17}$ As $s$ approaches $s_{0}$ from above along the real axis, $\Psi(\beta,z;s)$ diverges to infinity as a simple pole.$^{18}$ From eq \ref{eq4.1} the thermodynamics are then fully determined by
  \begin{equation}
    \beta p = s_{0}(\beta,z).  \label{eq4.3}
  \end{equation}

For a linear system with only {\em nearest-neighbor interactions}, one finds
  \begin{equation}
   \Psi(\beta,z;s) = z/[1-zJ(s)],  \label{eq4.4}
  \end{equation}
where $J(s)$ is the Laplace transform of the Boltzmann factor associated with the potential $\varphi(r)$, namely,
  \begin{equation}
    J(s)= \int_{0}^{\infty}e^{-sr}\exp[-\beta\varphi(r)]dr.  \label{eq4.5}
  \end{equation}
The abscissa of convergence, $s_{0}$, is determined by the simple pole in eq \ref{eq4.4}. From eq \ref{eq4.3} the equation of state may then be derived from$^{17}$
  \begin{equation}
   1/z = J(\beta p), \hspace{0.3in} -1/\rho z = J^{\prime}(\beta p),  \label{eq4.6}
  \end{equation}
where $J^{\prime}(s)$ denotes the first derivative of $J(s)$ with respect to $s$. 

For a discretized system the integral in eq \ref{eq4.2} must be replaced by a sum and, similarly, $J(s)$ in eqs \ref{eq4.4}-\ref{eq4.6} must be defined as
 \begin{equation}
  J(s;\zeta)=\sum_{l=0}^{\infty}e^{-sla_{0}}\exp[-\beta \varphi(la_{0})]a_{0},
 \end{equation}
where the identification $r=la_{0}$ with $l=0,1,2,\cdots$ has been adopted.

A one-dimensional model with finite-range potentials, $\varphi(r)$, such as we will consider, does not, of course, exhibit any phase separation or criticality. Nonetheless, questions regarding the critical point parameters, $T_{\mbox{\scriptsize c}}$, $\rho_{\mbox{\scriptsize c}}$, etc., can be approached$^{17}$ by introducing an appropriate van der Waals limit.$^{19}$ In this limit the equation of state may be written
  \begin{equation}
   p = p_{0}(\rho,T) - \epsilon a \rho^{2},  \label{eq4.7}
  \end{equation}
where $p_{0}(\rho,T)$ is the solution of eq \ref{eq4.6} while the second term, $\epsilon a \rho^{2}$, is chosen so as to match the exact high-temperature behavior of the second virial coefficient which results when an infinitely weak, infinitely long-range attractive interaction is added. In the following we investigate various short-range interaction potentials, $\varphi(r)$, and study the approach of the critical parameters for discretized systems to their continuum limits when $\zeta\rightarrow\infty$.

\subsection{Pure hard-core model}
\indent

The simplest model is the hard-core model with interaction potential
 \begin{eqnarray}
   \varphi(r) & = & \infty, \hspace{0.3in}  \mbox{for  $~~r<a$,} \nonumber \\
	      & = & 0, \hspace{0.37in}\mbox{for  $~~r\geq a$}.  
 \end{eqnarray}
In the continuum limit we obtain from eq \ref{eq4.6} the well known result
 \begin{equation}
   \beta p_{0}(\rho,T) = \rho/(1-a\rho), \label{eq4.2.2}
 \end{equation}
which in eq \ref{eq4.7} yields the standard vdW equation. From this, using the normal classical criticality conditions
 \begin{equation}
   (\partial p/\partial\rho)_{T}=0,\hspace{0.2in} (\partial^{2} p/\partial\rho^{2})_{T}=0, \hspace{0.2in} (\partial^{3} p/\partial\rho^{3})_{T}\neq 0,  \label{eq4.2.4a}
 \end{equation}
we find the usual results
 \begin{equation}
   \rho^{\ast}_{\mbox{\scriptsize c}}(\infty)\equiv a\rho_{\mbox{\scriptsize c}}(\infty)=\mbox{$\frac{1}{3}$}, \hspace{0.2in} T^{\ast}_{\mbox{\scriptsize c}}(\infty)\equiv k_{\mbox{\scriptsize B}}T_{\mbox{\scriptsize c}}(\infty)/\epsilon = \mbox{$\frac{8}{27}$},  \label{eq4.2.3}
 \end{equation}
while $p^{\ast}_{\mbox{\scriptsize c}}(\infty)\equiv ap_{\mbox{\scriptsize c}}(\infty)/\epsilon = \frac{1}{27}$.

Now, consider the discretized system with parameter $\zeta = a/a_{0}$. Let us first address the case in which $\zeta$ is an integer, say $n$ (as has been standard in the literature$^{2-8}$). It is then convenient to define an ``effective'' or lattice hard-core diameter $b_{0}\equiv na_{0}$. When $\zeta=n$ we have, of course, $b_{0}=a$. From eq \ref{eq4.6} we then easily obtain
 \begin{equation}
  \frac{a_{0}p_{0}(\rho,T;n)}{k_{\mbox{\scriptsize B}}T} = \ln\left(\frac{1-(b_{0}-a_{0})\rho}{1-b_{0}\rho}\right), \hspace{0.2in} b_{0}=na_{0}=a.  \label{eq4.2.4} 
 \end{equation}
From this, eqs \ref{eq4.7} and \ref{eq4.2.4a} the critical parameters are found to be
 \begin{equation}
  \rho_{\mbox{\scriptsize c}}^{\ast}(n) = b_{0}\rho_{\mbox{\scriptsize c}} = \frac{2n-1 - (n^{2}-n+1)^{1/2}}{3(n-1)}, \label{eq4.2.5}
 \end{equation}
 \begin{equation}
  T_{\mbox{\scriptsize c}}^{\ast}(n) = \frac{k_{\mbox{\scriptsize B}}T_{\mbox{\scriptsize c}}}{\epsilon} = 2a\rho_{\mbox{\scriptsize c}}(1-b_{0}\rho_{\mbox{\scriptsize c}})(1-b_{0}\rho_{\mbox{\scriptsize c}}+b_{0}\rho_{\mbox{\scriptsize c}}/n). \label{eq4.2.6}
 \end{equation}
When $n\rightarrow\infty$ the continuum values in eq \ref{eq4.2.3} are reproduced. By expanding in inverse powers of $n$, we find that the rate of convergence is described by
 \begin{equation}
   \frac{\Delta \rho_{\mbox{\scriptsize c}}(n)}{\rho_{\mbox{\scriptsize c}}(\infty)} \equiv \frac{[\rho_{\mbox{\scriptsize c}}(n)-\rho_{\mbox{\scriptsize c}}(\infty)]}{\rho_{\mbox{\scriptsize c}}(\infty)} = \frac{1}{2n}\left(1+\frac{1}{4n} - \frac{1}{8n^{2}} + \cdots \right),  \label{eq4.2.7}
 \end{equation}
 \begin{equation}
   \frac{\Delta T_{\mbox{\scriptsize c}}(n)}{T_{\mbox{\scriptsize c}}(\infty)} \equiv \frac{ [T_{\mbox{\scriptsize c}}(n)-T_{\mbox{\scriptsize c}}(\infty)]}{T_{\mbox{\scriptsize c}}(\infty)} = \frac{1}{2n}\left( 1 + \frac{3}{8n} + \frac{1}{16n^{2}} + \cdots \right).  \label{eq4.2.8}
 \end{equation}
As anticipated from the analysis of the hard-core part of the second virial coefficient presented in the previous section, $\rho_{\mbox{\scriptsize c}}(n)$ and $T_{\mbox{\scriptsize c}}(n)$ both converge to their limiting values as $n^{-1}$, i.e., with exponent $\psi=1$.

Apart from matters of convenience or tradition$^{10}$ there are no reasons for restricting the choice of $\zeta$ to integers. However, one may reasonably guess that the truncation errors in the critical parameters will still decay as $1/\zeta$ for most values of $\zeta$. But an interesting question arises: Are there any particular values of $\zeta$ that improve the convergence rates? In other terms, if we put $\zeta=n-\delta$ with $0\leq\delta <1$, are there $\delta$ values such that $\rho_{\mbox{\scriptsize c}}(\zeta)$ and $T_{\mbox{\scriptsize c}}(\zeta)$ converge to their limits faster than $n^{-1}$?

To address this question, consider the second virial coefficient, $B_{2}(T)$, for the potential $\varphi(r)$, in both the continuum and the lattice: see eqs \ref{eq.3} and \ref{eq.4} for $d=1$. Certainly, one has $B_{2}^{\infty}(T)=a$ in the continuum while $B_{2}^{\zeta}=(n-\frac{1}{2})a_{0}$ in the discrete space when $\zeta=n-\delta$. Note that, by definition, $a_{0}=a/\zeta=a/(n-\delta)$. It is then plausible that the choice of $\delta$ which matches $B_{2}^{\zeta}$ to $B_{2}^{\infty}$ may improve the convergence of $\rho_{\mbox{\scriptsize c}}(\zeta)$ and $T_{\mbox{\scriptsize c}}(\zeta)$. In the hard-core model, this value is simply $\delta=\frac{1}{2}$.

Accordingly, let us repeat the previous calculations but with $\zeta=n-\frac{1}{2}$. But the hard-core part of the pressure, $p_{0}(\rho,T)$, given in eq \ref{eq4.2.4} does not change! Thus the same results, eqs \ref{eq4.2.5} and \ref{eq4.2.6}, for $\rho_{\mbox{\scriptsize c}}$ and $T_{\mbox{\scriptsize c}}$ still apply. Note, however, that one now has $a_{0}=a/(n-\frac{1}{2})$ and $b_{0}=na/(n-\frac{1}{2})$. Substituting these into eqs \ref{eq4.2.5} and \ref{eq4.2.6} yields
 \begin{equation}
  \frac{\Delta\rho_{\mbox{\scriptsize c}}(\zeta)}{\rho_{\mbox{\scriptsize c}}(\infty)} = -\frac{1}{8n^{2}}\left( 1 +\frac{1}{n}+\frac{7}{16n^{2}} + \cdots \right),  \label{eq4.2.10} 
 \end{equation}
 \begin{equation}
  \frac{\Delta T_{\mbox{\scriptsize c}}(\zeta)}{T_{\mbox{\scriptsize c}}(\infty)} = -\frac{1}{16n^{2}}\left(1 + \frac{1}{n} +\frac{9}{32n^{2}} + \cdots\right).  \label{eq4.2.11}
 \end{equation}

Evidently the hoped-for improvement in convergence has been realized! Indeed, not only does the choice $\delta$$\,=\,$$\frac{1}{2}$ yield faster convergence by a factor $1/n$$\,\sim\,$$ 1/\zeta$, but the magnitudes of the leading amplitudes here are much smaller than those in eqs \ref{eq4.2.7} and \ref{eq4.2.8}. It is also worth noting that they are of opposite sign. Consequently we learn that convergence from above or from below should not be a universal feature.

\subsection{Square-well models}
\indent

To explore further let us consider hard-core square-well models of well depth $\epsilon >0$. The interaction potential is thus
 \begin{eqnarray}
  \varphi( r) & = & \infty, \hspace{0.3in} \mbox{for $~~r<a$}, \nonumber \\
              & = & -\epsilon, \hspace{0.26in} \mbox{for $~~a\leq r < c=\lambda a$}, \nonumber \\
              & = & 0, \hspace{0.38in} \mbox{for $~~r\geq c$},
 \end{eqnarray}
with $\lambda >1$. Realistic values satisfy $\lambda < 2$ but, for the present purposes we may also examine $\lambda \geq 2$.

As in the previous section, let us first suppose that $\zeta = n$ is an integer; but since $\lambda$ need not be integral it is helpful to employ the (slightly nonstandard) notation $[x]=\mbox{min}_{m}\{ m\geq x\}$ where $m$ is an integer, so that $[x]$ is the smallest integer not less than $x$. Then the required Laplace transform is
 \begin{equation}
  J(s;\zeta) = a_{0} \frac{e^{\beta\epsilon - na_{0}s} + (1-e^{\beta\epsilon})e^{-[\lambda n]a_{0}s}}{1-e^{-a_{0}s}}, 
 \end{equation}
where, once more, $a_{0}=a/\zeta = a/n$. The pressure $p_{0}(\rho,T)$ again follows from the relations eq \ref{eq4.6} but an analytical treatment is no longer tractable. Nonetheless, by solving the critical point conditions eq \ref{eq4.2.4a} numerically, precise values for $T_{\mbox{\scriptsize c}}$ and $\rho_{\mbox{\scriptsize c}}$ are readily found. The results for $\lambda=1\frac{1}{2}$, $2$, $3$ and $5$ are presented in Figure 4. As expected from the previous analysis of the pure hard-core model, $\rho_{\mbox{\scriptsize c}}(n)$ and $T_{\mbox{\scriptsize c}}(n)$ converge to their limits as $1/n$.

Now the previous question again arises, namely, can one find nonintegral values of $\zeta$ that provide enhanced rates of convergence? It is reasonable to anticipate that if such values exist, the corresponding second virial coefficients $B_{2}^{\zeta}(T)$ [for the potential $\varphi( r)$] will also converge faster to the limit $B_{2}^{\infty}(T)$. Accordingly, let us again look for {\em non}$\,$integral values of $\zeta=a/a_{0}$ for which $B_{2}^{\zeta}(T)=B_{2}^{\infty}(T)$. To simplify the formulae it is convenient to write $c/a_{0}=\theta$ and to suppose (as will transpire) that $\theta$ is also nonintegral. Then we find
 \begin{eqnarray}
  B_{2}^{\zeta}(T) & = & ([\zeta] -\mbox{$\frac{1}{2}$})a_{0} + ([\theta]-[\zeta])a_{0}(1-e^{-\beta\epsilon}), \nonumber \\
  & \rightarrow & a + (c-a)(1-e^{-\beta\epsilon}), \hspace{0.35in} \mbox{when  $~~\zeta\rightarrow\infty$.}  \label{eq3.B2}
 \end{eqnarray}

It is possible that there are values of $\zeta$ that solve the equality $B_{2}^{\zeta}(T)=B_{2}^{\infty}(T)$ and depend explicitly on $\beta\propto 1/T$. However, such $T$-dependent choices for $\zeta$ are hardly acceptable when the actual aim is to determine the unknown value of $T_{\mbox{\scriptsize c}}$. Accordingly, we will seek only $T$-independent solutions. To that end we may first let $\beta\rightarrow 0$: in this limit the second terms on the right hand sides of eq \ref{eq3.B2} vanish and the model reduces to the pure hard-core case. Equating the first terms in eq \ref{eq3.B2} yields $\zeta = [\zeta] -\frac{1}{2}$ which is generally solved by $\zeta = n-\frac{1}{2}$ with $n=1,2,3,\cdots\:$. Of course, this just confirms the previous conclusion. Equating the second terms (for $n<\infty$ and for $n=\infty$) similarly yields $\theta = [\theta] - \frac{1}{2}$ so that one must also have $\theta = m - \frac{1}{2}$ with $m=1,2,3,\cdots\:$.

But at this point one must observe that 
 \begin{equation}
  \lambda = \frac{c}{a}=\frac{\theta}{\zeta}=\frac{m-\frac{1}{2}}{n-\frac{1}{2}} = \frac{2m-1}{2n-1}, \label{eq.lambda}
 \end{equation}
is to be held {\em fixed} when one allows $\zeta = n-\frac{1}{2}$ to diverge. Evidently this is possible {\em only if} $\lambda$ is the ratio of two {\em odd integers} say, in lowest terms, $\lambda = (2j+1)/(2k+1)$ with $j$ and $k$ nonnegative integers and $j>k$ (to preserve $\lambda >1$). If numerator and denominator here are multiplied by $(2l+1)$, where $l$ is a positive integer, the odd/odd character of $\lambda$ is preserved. Comparing with eq \ref{eq.lambda} we then quickly find that the arithmetical sequences
 \begin{equation}
  \zeta = \zeta_{k}(n) = (2k+1)(n-\mbox{$\frac{1}{2}$}) \hspace{0.3in} (n=1,2,3,\cdots)
 \end{equation}
satisfy the second virial coefficient equality provided $\lambda$ takes a value of the form $(2j+1)/$ $(2k+1)$.

In the simplest case, $k=0$, the ratio $\lambda = c/a$ can only be an odd integer. When $k=1$ further acceptable or ``conforming'' values are $\lambda=1\frac{2}{3}$, $2\frac{1}{3}$, $3\frac{2}{3}$, $\cdots$, with $\zeta = 4\frac{1}{2}$, $7\frac{1}{2}$, $10\frac{1}{2}$, $\cdots\:$. Likewise, $k=2$ yields $\lambda = \frac{7}{5}$, $\frac{9}{5}$, $\frac{11}{5}$, $\cdots\;$ and $\;\zeta= 7\frac{1}{2}$, $12\frac{1}{2}$, $\cdots\:$. Figure 5 presents critical densities and temperatures vs $1/\zeta^{2}$ for the optimal sequences $\zeta=\zeta_{k}$ for the models with $\lambda=1\frac{2}{3}$, $2\frac{1}{3}$, $3$ and $5$. The expected enhanced convergence with exponent $\psi = 2$ is clearly confirmed for these special, conforming values of $\lambda$. Notice, furthermore, that even for relatively small values of $\zeta$, the magnitudes of the truncation errors in $T_{\mbox{\scriptsize c}}$ and $\rho_{\mbox{\scriptsize c}}$ are ten-fold smaller than for the integral $\zeta$ values. Conversely, if $\lambda$ is not the ratio of two odd integers one cannot increase the convergence exponent merely by choosing some optimal arithmetical sequence of $\zeta$ values. Nonetheless, when $\lambda$ is close to a conforming value the related $\zeta_{k}$ values should yield more accurate estimate for $\rho_{\mbox{\scriptsize c}}(\infty)$ and $T_{\mbox{\scriptsize c}}(\infty)$.

Finally, we must point out that despite the somewhat tortuous derivation of the conforming $\lambda$ values and the corresponding optimal sequences $\zeta_{k}(n)$, there really is no mystery in these results! Rather one need only focus on the dominant feature of the hard-core square-well models, namely that the Mayer $f$-functions or, equivalently, the Boltzmann factors $e^{-\beta\varphi(r)}$, exhibit discontinuities at $r=a$ (as for the pure hard-core model) and at $r=c$: see Figure 6(a). The slow $1/\zeta$ decay of the truncation errors is simply ordained by these discontinuities. 

To see this clearly in the general case of approximating a single-variable integral by a simple sum over lattice sites spaced at intervals $r_{l+1}-r_{l} = a_{0}$, suppose that the integrand has a discontinuity of magnitude $\Delta$ at $r=b$. If the closest lattice sites are located at $r_{+}= b+a_{0}\delta$ and $r_{-}=b-a_{0}\delta^{\prime}$ with $(\delta+\delta^{\prime})=1$ and $\delta,\delta^{\prime}<1$, the leading error in approximating the integral through $r=b$ by a sum is of order $\Delta (\frac{1}{2}-\delta)a_{0}\propto \Delta a/\zeta$: this is easily seen graphically. However, if $r=b$ lies {\em midway} between adjacent lattice sites, so that $\delta=\delta^{\prime}=\frac{1}{2}$, the leading error term, proportional to $\Delta$, vanishes identically! Instead, if the integrand has bounded derivatives near $r=b$, the error is only of order $a_{0}^{2}= a^{2}/\zeta^{2}$. 

When there are two discontinuities in the integrand (and the origin is fixed at a lattice site) an essentially number theoretic problem arises: namely, one must find a lattice spacing $a_{0}$ so that both points of discontinuity at $r=a$ and $r=c$ lie midway between adjacent lattice points. From this perspective the condition $\zeta=a/a_{0}=n-\frac{1}{2}$ and the restriction eq \ref{eq.lambda} on $\lambda$ are rather obvious. Ameliorating further discontinuities would require further restrictions on their relative locations.

We conclude from this discussion that if the Mayer function, $f(r)$, is continuous (but perhaps with discontinuities in its derivatives), the critical parameters should converge more rapidly than $1/\zeta$. To investigate this issue we examine first a model that can display a logarithmically softened repulsive core.

\subsection{Logarithmic repulsive core models}
\indent

We define the logarithmic repulsive core potential by
 \begin{eqnarray}
  \varphi(r) & = & \infty, \hspace{1.5in} \mbox{for $r<a$}, \nonumber \\
             & = & -\epsilon\ln \left( w + w^{\prime}\frac{r-a}{c-a}\right), \hspace{0.13in} \mbox{for $a\leq r < c =\lambda a$}, \nonumber \\
             & = & 0, \hspace{1.58in} \mbox{for $r\geq c$},
 \end{eqnarray}
where the parameter $w=1-w^{\prime}$ controls the continuity of $e^{-\beta\varphi(r)}$ at $r=a$. Indeed, when $w=0$ the potential $\varphi(r)$ diverges smoothly to $+\infty$ as $r\rightarrow a+$; however, for $0<w<1$ the potential rises only to $\epsilon\ln(1/w)<\infty$. Thus the Mayer $f$-function is continuous everywhere when $w=0$ while it exhibits a discontinuity at $r=a$ whenever $w\neq 0$: see Figure 6(a). Furthermore, the derivative, $df(r)/dr$, is discontinuous at $r=c$ for all values of $w$. One may then anticipate that for $w\neq 0$ both $\rho_{\mbox{\scriptsize c}}(n)$ and $T_{\mbox{\scriptsize c}}(n)$ converge to their limiting values as $n^{-1}$ while for $w=0$ the convergence should be faster.

To confirm this, we consider for simplicity only integer values of $\zeta$ and set $\lambda=2$. The Laplace transform of the potential in the discretized system is then
 \begin{equation}
  J(s;\zeta) = a_{0}\frac{e^{-2na_{0}s}}{1-e^{-a_{0}s}} + a_{0}\sum_{l=n}^{2n-1} e^{-la_{0}s}\left( w+w^{\prime}\frac{l-n}{n}\right)^{\beta\epsilon}, \label{eq4.4.2}
 \end{equation}
where $a_{0}=a/n$. From this we are able to calculate $p_{0}(\rho,T)$ numerically, obtain the equation of state, and solve to obtain the critical point values illustrated in Figure 7 for different values of $w$. The critical parameters for $w=0$ clearly converge to the continuum limits as $1/n^{2}$, faster than the standard $1/n$ rate observed for the previous models with ``abrupt hard cores'': see the insets in Figure 7. On the other hand, for $w\neq 0$ the truncation errors decrease, as expected, only like $1/n$.

As explained, the slow $1/n$ decay when $w\neq 0$ arises from the hard-core discontinuity of $f(r)$ at $r=a$. By choosing $\zeta=n-\frac{1}{2}$ so that the hard-core part of the second virial coefficient for the discretized system matches the continuum counterpart, we again expect to obtain faster convergence. Indeed, as illustrated in Figure 8, those values of $\zeta$ do lead to a $1/\zeta^{2}$ rate of convergence of $\rho_{\mbox{\scriptsize c}}(\zeta)$ and $T_{\mbox{\scriptsize c}}(\zeta)$ to their limits.

\subsection{Cubic model}
\indent

As a last case it is natural to consider a potential, $\varphi(r)$, smoother than the previous examples and ask how rapidly the critical-point values approach their continuum limits: We expect to find a rate faster than $1/n^{2}$. For this purpose, consider a "cubic model" defined by the potential
 \begin{eqnarray}
  \varphi(r) & = & \infty, \hspace{2.9in} \mbox{for $r<a$}, \nonumber \\
		 & = & -\epsilon\ln\hspace{-0.04in} \left[\mbox{$\frac{1}{2}$}+w_{1}\left(\frac{r}{a}-\lambda_{1}\right)-w_{2}\left(\frac{r}{a}-\lambda_{2}\right)^{\Theta}\right],\hspace{0.1in} \mbox{for $a\leq r <c$}, \nonumber \\
		 & = & 0,\hspace{2.98in} \mbox{for $r\geq c$},
 \end{eqnarray}
where $\Theta$, $w_{1}$, $w_{2}$, and $\lambda_{1}$, and $\lambda_{2}$ are parameters. For illustration and to ensure $\varphi(r)$ and $(d\varphi/dr)$ are continuous at the boundaries, $r=a$ and $c=\lambda a$, we choose $\Theta = 3$, $\lambda = 2$, $\lambda_{1}=\lambda_{2}=\frac{3}{2}$, $w_{1}=\frac{3}{2}$ and $w_{2}=2$. With these choices, $f(r)$ and $(df/dr)$ become continuous everywhere: see Figure 6(b). 

The second virial coefficient for this potential with $\zeta=n$ is given by
  \begin{equation}
   B_{2}^{n}(T) = a_{0}\left[n-\mbox{$\frac{1}{2}$} - 2\sum_{l=n}^{2n-1}(e^{-\beta\varphi(la_{0})}-1) \right],  \label{eq4.5.2}
 \end{equation}
with $a_{0}=a/n$. This can be calculated numerically for any temperature. It is then easily seen that $B_{2}^{n}(T)$ converges to $B_{2}^{\infty}(T)$ as $1/n^{4}$: this is just as anticipated since $f(r)$ and its first derivative are continuous.

The critical point for this model can also be computed numerically via eqs \ref{eq4.6} and \ref{eq4.7} by using the Laplace transform
 \begin{equation}
  J(s) = a_{0}\frac{(1-e^{\beta\epsilon})e^{-2na_{0}s}}{1-e^{-a_{0}s}} +a_{0} \sum_{l=n}^{2n-1}e^{-la_{0}s}\left[\frac{1}{2}+\frac{3}{2}\left(\frac{l}{n}-\frac{3}{2}\right)-2\left(\frac{l}{n}-\frac{3}{2}\right)^{3}\right]^{\beta\epsilon}.  \label{eq4.5.3}
 \end{equation}
The results are presented in Figure 9. We observe that both $\rho_{\mbox{\scriptsize c}}(n)$ and $T_{\mbox{\scriptsize c}}(n)$ converge to the limiting values as $1/n^{4}$ just as indicated by the behavior of the second virial coefficient.

\section{Extensions for Higher Dimensions}
\indent

So far we have discussed in explicit and analytical detail the effects of the discretization parameter, $\zeta$, on the critical temperature and density, $T_{\mbox{\scriptsize c}}$ and $\rho_{\mbox{\scriptsize c}}$, of various one-dimensional models. A central lesson that we have learned from the exact calculations is that the behavior of the critical temperature and density in discretized systems closely reflects the convergence of the discretized second virial coefficient, $B_2^\zeta (T)$. As indicated in the Introduction and discussed further in Appendix A, one may extend this notion to analyze more realistic, higher dimensional systems, e.g., two- or three-dimensional model fluids. As we will show via explicit numerical calculations and simulations, the basic conclusion is confirmed. We aim here, in particular, to discover techniques for estimating the limiting critical points more precisely on the basis of less computationally expensive simulations of the discretized versions.

\subsection{Smooth potentials}
\indent

With this end in mind, let us first examine the convergence behavior of the discretized second virial coefficients, for typical smooth potentials $\varphi(r)$ in $d=3$ dimensions. It is convenient to normalize via
 \begin{equation}
  B_{2}^{\ast}(T;\zeta) \equiv B_{2}^{\zeta}(T)/(2\pi\sigma^{3}/3),
 \end{equation}
where $\sigma\:(\equiv a)$ is the diameter of the repulsive core of the potential defined for single-well smooth potentials, as usual, by $\varphi(\sigma)=0$. Of prime interest is the Lennard-Jones (LJ) $(12,6)$ potential,$^{20}$ with well-depth $\epsilon$, for which the original discretized simulations$^{3}$ with $\zeta=1$, $2$, $3$, $5$ and $10$ indicated rather satisfactory agreement with well established continuum estimates of $T_{\mbox{\scriptsize c}}$. However, these initial studies did not examine carefully the behavior of $T_{\mbox{\scriptsize c}}(\zeta)$ as a function of $\zeta$. Subsequent investigations$^{5}$ employed the modified Buckingham exponential-6 (E-6) potential$^{21,22}$ and looked systematically at the dependence of $T_{\mbox{\scriptsize c}}(\zeta)$ and $\rho_{\mbox{\scriptsize c}}(\zeta)$ on $\zeta$ for the values $\zeta=5$, $6$, $\cdots\:$, $10$ and $\zeta=15$. Accordingly, we will also examine $B_{2}^{\ast}(T;\zeta)$ for the modified E-6 potential.

The Boltzmann factors, $e^{-\beta\varphi(r)}$, for the two models at the estimated critical temperatures, $k_{\mbox{\scriptsize B}}T_{\mbox{\scriptsize c}}/\epsilon=1.299$ and $1.243$, respectively$^{3,23}$ are presented in Figure 10. For the LJ potential the Mayer $f$-function is smooth everywhere, all its derivatives being continuous. On the other hand, the modified E-6 potential exhibits a discontinuity at $r_{\mbox{\scriptsize max}}\simeq 0.2285\sigma$ owing to the hard-core cutoff.$^{22}$ However, the discontinuity is invisible in this figure and irrelevant in the computations since its magnitude is less than $10^{-9637}$.

Figure 11 presents the reduced, discretized second virial coefficients for the LJ and modified E-6 potentials evaluated for integral values of $\zeta\leq 25$ at the best estimates of the corresponding critical points. Since, of course, $B_{2}^{\ast}(T;\zeta)$ is an analytic function of $T$ for all $\zeta\leq\infty$, the precise value of $T$ should have a rather small influence on the $\zeta$ variation. As anticipated from the smoothness of the LJ and modified E-6 potentials (see Figure 10), both second virial coefficients converge rapidly to their continuum limits. (The latter are explicitly marked on the figures; needless to say, the continuum second virial coefficients for such radially symmetric potentials can be computed to high accuracy using standard integration routines.) This figure demonstrates that over a computationally relevant range of discretization parameters (i.e., $\zeta\lesssim 20$) the observed rates of convergence are well described by a $1/\zeta^{6}$ dependence. Indeed, as mentioned in the Introduction, an effective convergence exponent $\psi\simeq 6$ has been observed numerically in the variation of $T_{\mbox{\scriptsize c}}(\zeta)$ for the modified E-6 potential.$^{5}$ (For this model, to within the precision of the simulations, no significant variation of $\rho_{\mbox{\scriptsize c}}(\zeta)$ could be resolved.$^{5}$)

Although the agreement between the observed convergence behavior of $T_{\mbox{\scriptsize c}}(\zeta)$ and the second virial coefficient, $B_{2}^{\zeta}$, in a more-or-less realistic three-dimensional model is gratifying, the value $\psi\simeq 6$ should not be taken too seriously! Thus, it would be interesting to see how far the same effective rate of convergence ($\psi \simeq 6$) is maintained for both models when the evaluation of the discretized second virial coefficients for the LJ and modified E-6 potentials is pushed further (and, hence, beyond any normally relevant level of precision). However, significant further efforts are needed to resolve the issue numerically and we have not undertaken the task. Of course, for extremely large values of $\zeta$ we expect a $\sqrt{\ln\zeta}/\zeta^{2}$ asymptotic convergence law for the modified E-6 potential to appear in view of the sharp, although exceedingly small discontinuity in $f(r)$.

\subsection{Potentials with hard cores}
\indent

Because of their conceptual simplicity and theoretical interest, it is important to address models where $\varphi(r)$ consists of a hard-core potential $\varphi_{0}(r)$ plus a bounded and smooth attractive potential $\varphi_{1}(r)$. Such systems include both the RPM and the hard-core dumbbell models studied in ref 5. Certainly the convergence of the second virial coefficients will be dominated by the behavior of the hard-core part $B_{0}^{\zeta}$ (which, of course, is $T$-independent). Thus, following the arguments for $d=1$, we may expect that the critical temperatures and densities will approach their continuum limits as $\zeta\rightarrow\infty$ in a manner reflecting the variation of $B_{0}^{\zeta}$.

As seen in section II the hard-core truncation error $E_{0}(T)\equiv (B_{0}^{\zeta}/B_{0}^{\infty})-1$ decays as $1/\zeta^{2}$ for all $d\geq 4$ while for $d=3$ the error varies as $(\ln\zeta)^{1/2}/\zeta^{2}$. However, as demonstrated in Figure 2(b) (for $d=2$) and as shown in Figure 3 of ref 8 (for $d=3$), the behavior of the truncation error for $d\geq 2$ is intrinsically erratic, chaotic or `noisy'. Indeed, in $d=3$ dimensions it may be viewed as a random variable with, as discussed, a Gaussian distribution: recall Figure 3. Furthermore, it is found in the simulations that the estimates of $T_{\mbox{\scriptsize c}}(\zeta)$ and $\rho_{\mbox{\scriptsize c}}(\zeta)$ for, e.g.,$^{8}$ the RPM, also exhibits noisy behavior that, at least superficially, is reminiscent of that which necessarily arises in $B_{0}^{\zeta}$.

To proceed we might initially ask, as in section III, if one might not be able to select special sequences of values for $\zeta$ (not necessarily integers) so that $T_{\mbox{\scriptsize c}}(\zeta)$ and $\rho_{\mbox{\scriptsize c}}(\zeta)$ would converge faster than otherwise. To that end one should first ask (in $d\geq 2$ dimensions) for sequences of $\zeta$ values for which $B_{0}^{\zeta}=B_{0}^{\infty}$ so that $E_{0}(\zeta)$ vanishes identically. Such {\em matching sequences} would seem good candidates for ensuring the optimally rapid convergence of $T_{\mbox{\scriptsize c}}(\zeta)$ and $\rho_{\mbox{\scriptsize c}}(\zeta)$ as verified in section III for $d=1$: see Figures 5 and 8.

Inspection of Figure 2(b) and Figure 3 in ref 8 leaves little doubt that such infinite matching sequences exist for $d=2$ and $3$. But one is easily convinced that integral values of $\zeta$ cannot appear; further, one may seriously doubt that even arithmetical sequences like $kn-\delta$ with fixed $\delta$ and integer $k$ could satisfy the matching condition. However, there is no obstacle to the computation of matching values to any required precision up to relevantly large values of $\zeta$. Explicitly, in Table I we present, for both $d=2$ and $3$, {\em selected} matching values of $\zeta$ in the range $(5,25)$. In planning simulations of models with hard cores it seems likely to be advantageous to choose such matching values. However we have not undertaken simulations to verify this surmise.

It maybe worth mentioning that since periodic boundary conditions$^{10}$ for a range of box sizes must be employed (in order that essential finite-size extrapolation techniques can be applied systematically$^{11}$) a purely cosmetic blemish will arise if matching values of $\zeta$ are adopted: specifically, since $L^{\ast}\zeta$ must be an integer (where $L^{\ast}\equiv L/a$ is the reduced linear dimension of the simulation box),$^{10}$ nonintegral and, indeed, irrational values of $L^{\ast}$ will be entailed. However, this feature is of no consequence in any of the computations entailed in implementing the fine-discretization technique,$^{2-4}$ neither setting-up, data gathering nor analysis.$^{11}$

However, whether or not matching $\zeta$-values are employed for hard-core systems (and whether or not they prove efficacious to some degree in practice) it is reasonable to ask how the difficulties of extrapolation `through inherent residual truncation noise' might be ameliorated. Explicitly, faced with evidently erratic estimates for $T_{\mbox{\scriptsize c}}(\zeta)$ and $\rho_{\mbox{\scriptsize c}}(\zeta)$ such as found for the RPM in ref 8 (or for hard-core dumbbell models in ref 5), how might reliable estimates for the continuum parameters be obtained?

Following ref 8 we describe two approaches. First, noting that the general behavior of $T_{\mbox{\scriptsize c}}(\zeta)$ and $\rho_{\mbox{\scriptsize c}}(\zeta)$ resembles that of $B_{0}^{\zeta}$ and accepting the idea developed in section III that this should be more than coincidental, let us attempt matching {\em post facto}. Thus we may introduce a modified or {\em smoothed discretization level}, $\zeta^{\dag}(\zeta)$ defined so that, for $d=3$,
 \begin{equation}
   E_{0}(\zeta) = c^{\dag}/\zeta^{\dag\,2}, \label{eq.E0}
 \end{equation}
where $c^{\dag}$ is an assigned constant. (Here we neglect, for the time being, the factor $(\ln\zeta)^{1/2}$ known to be present in the exact asymptotic behavior.) By construction, the truncation error $E_{0}(\zeta)$ now decays smoothly as $1/\zeta^{\dag\,2}$. We may thus hope that $T_{\mbox{\scriptsize c}}(\zeta)$ and $\rho_{\mbox{\scriptsize c}}(\zeta)$ will also behave smoothly when plotted vs $1/[\zeta^{\dag}(\zeta)]^{2}$. In order for this method to be convincing, however, the resulting extrapolated estimates should not depend significantly on the choice of $c^{\dagger}$. To demonstrate that this is indeed so for the RPM, we present in Figure 12 plots of $T_{\mbox{\scriptsize c}}(\zeta)$ and $\rho_{\mbox{\scriptsize c}}(\zeta)$ as functions of $\zeta^{\dag}(\zeta)$ computed with the choices $c^{\dagger}=15E_{0}(5)$, $25E_{0}(5)$, and $35E_{0}(5)$. [We remark that in ref 8 only $c^{\dagger}=25E_{0}(5)$ was used.] Evidently the behavior is fairly smooth and, in fact, much less erratic than plots vs.\ $1/\zeta^{2}$: see Figure 4(i) in ref 8. From these plots we conclude $T_{\mbox{\scriptsize c}}^{\ast}(\infty)\equiv k_{\mbox{\scriptsize B}}T_{\mbox{\scriptsize c}}Da/q_{0}^{2}=0.04933(5)$ and $\rho_{\mbox{\scriptsize c}}^{\ast}(\infty)\equiv \rho_{\mbox{\scriptsize c}}a^{3}=0.0750(10)$ for the continuum RPM.$^{8}$

Of course, the presence of the logarithmic factor in the true asymptotic decay of $E_{0}(\zeta)$ raises a question regarding the validity of this smoothing procedure. However, one may anticipate that the logarithmic factor plays an insignificant role for the fairly small values of $\zeta=5$-$20$ employed in these simulations of the RPM. Indeed, one may readily introduce a factor $(\ln\zeta)^{1/2}$ in the definition of $\zeta^{\dag}(\zeta)$ in eq \ref{eq.E0}; but on doing so we find that to within the simulation uncertainties the approach provides the same estimates for $T_{\mbox{\scriptsize c}}(\infty)$ and $\rho_{\mbox{\scriptsize c}}(\infty)$. It goes without saying, however, that the effectiveness of this method for the RPM is not much more than a promise of success in other cases; but reasonable optimism seems in order!

\subsection{Rescaling to accelerate convergence}
\indent

Another approach to improving convergence that we have explored to a limited degree is based on the idea that what is needed in obtaining optimal estimates from a fine-discretization calculation is some appropriate {\em rescaling} of density and temperature depending on $\zeta$. Our analysis again suggests that this might profitably be guided by the behavior of the {\em full} discretized second virial coefficient,
 \begin{equation}
  B_{2}^{\zeta}(T) = B_{0}^{\zeta}(T) + B_{1}^{\zeta}(T),  \label{eq.B2total}
 \end{equation}
rather than just the repulsive part $B_{0}^{\zeta}(T)$. Indeed, the contribution of the attractive tail $\varphi_{1}(r)$ must surely play a role and a hard core might be softened as in the LJ and E-6 potentials. It is then natural to examine the behavior of the rescaled discretized critical densities defined via
 \begin{equation}
  \rho_{\mbox{\scriptsize c}}^{\dag}(\zeta) \equiv \rho_{\mbox{\scriptsize c}}(\zeta)B_{2}^{\zeta}(T)/B_{2}^{\infty}(T), \label{eq.rhoscale}
 \end{equation}
in the hope that they will vary more smoothly vs $1/\zeta^{\psi}$ for appropriate $\psi$.

To go a little further, one might, in as far as $\varphi_{1}(r)$ varies relatively smoothly by definition, approximate the full second virial coefficients in eq \ref{eq.rhoscale}, by replacing the contribution $B_{1}^{\zeta}(T)$ by a $\zeta$-independent value, say $b_{1}$. Indeed, when one considers the RPM, some modification of eq \ref{eq.rhoscale} is essential since, by virtue of the long-range Coulomb interactions, the first low density correction to ideal gas behavior is universal and proportional to $\rho^{3/2}$ rather than to $\rho^{2}$. And although a $\rho^{2}$ term also exists, it is dominated by a $\rho^{2}\ln\rho$ term.$^{24}$ Naively, one might be tempted to say that the second virial coefficient is divergent; but in reality although a hard-core term of the form $B_{0}(T)$ plays a part, there is no direct meaning that can be ascribed to $B_{1}(T)$. However, if $B_{1}^{\zeta}(T)$ in eq \ref{eq.B2total} is replaced by a parameter $b_{1}^{\rho}$, we may regard the density rescaling eq \ref{eq.rhoscale} merely as a semiphenomenological device. A similar expression, but with a parameter $b_{1}^{T}$, may then be used to rescale $T_{\mbox{\scriptsize c}}(\zeta)$ to obtain a $T_{\mbox{\scriptsize c}}^{\dag}(\zeta)$. Then the aim should be to find suitable choices of $b_{1}^{\rho}$ and $b_{1}^{T}$ that provide enhanced convergence.

This procedure has been successfully demonstrated for the RPM in ref 8. There it is seen (in Figure 4) that the choices $b_{1}^{\rho}=0.4B_{0}^{\infty}$ and $b_{1}^{T}=1.7B_{0}^{\infty}$ yield plots of $\rho_{\mbox{\scriptsize c}}^{\dag}(\zeta)$ and $T_{\mbox{\scriptsize c}}^{\dag}(\zeta)$ that display very little $\zeta$ dependence in the range $\zeta=5$-$20$ (whether plotted vs.\ $1/\zeta^{2}$, as seems most appropriate, or employing some other $\psi$ values). Indeed, extrapolation of $\rho_{\mbox{\scriptsize c}}^{\dag}(\zeta)$ and $T_{\mbox{\scriptsize c}}^{\dag}(\zeta)$ is then limited almost entirely by the Monte Carlo simulation uncertainties at the given $\zeta$ values: and the same estimates are obtained for the RPM as quoted above.

Finally, as the previous remark brings out, the practical application of the extrapolation methods just outlined to simulations of model fluids relies on the use of methods capable of providing precise and reliable estimates of $T_{\mbox{\scriptsize c}}$ and $\rho_{\mbox{\scriptsize c}}$ at a number of fixed discretization levels.$^{11}$ On the other hand, although the approach is as yet untested for $d=2$ and $3$, the use of the favored {\em matching discretization levels} listed in Table I may provide the best strategy if an investment in only one or two simulations at chosen $\zeta$ values can be justified.

\section{Summary}
\indent

In conclusion we have studied the effects of the lattice discretization level, $\zeta$, on the critical parameters, in particular, the critical temperature, $T_{\mbox{\scriptsize c}}$, and density, $\rho_{\mbox{\scriptsize c}}$, of model fluids. The lattice discretization technique can elevate the speed of simulations for complex fluids by factors of 10$\,$-100 relative to the corresponding continuum models. It has been demonstrated that the convergence of $T_{\mbox{\scriptsize c}}(\zeta)$ and $\rho_{\mbox{\scriptsize c}}(\zeta)$ as $\zeta$ increases may be understood rather straightforwardly by studying the discretized second virial coefficient, $B_{2}^{\zeta}(T)$. Appendix A discusses this conclusion from a general theoretical perspective. In particular, when the interaction potential contains a hard-core repulsive part, the partial second virial coefficient, $B_{0}^{\zeta}$, for the hard-core contribution (which is essentially given by the number of lattice points inside a $d$-dimensional sphere of radius $\zeta$) dominates the convergence of $B_{2}^{\zeta}(T)$. Rigorous aspects of this issue, namely how the number of lattice points converges to the volume of the sphere, were presented in section II. The relative truncation error in $B_{2}(T)$, namely $E_{0}(\zeta)$ as defined in eq \ref{eq.5}, was shown to decay as $c_{d}/\zeta^{(d+1)/2}$ for $d\leq 3$ with $c_{d}\sim O(1)$ for $d=1$ and $2$ but $c_{3}\sim \sqrt{\ln \zeta}$, while for all $d\geq 4$ one has $E_{0}(\zeta)\sim 1/\zeta^{2}$. These results can be understood intuitively on the basis of heuristic arguments presented in Appendix B.

To illustrate the general arguments in further detail, we have studied various one-dimensio-nal model fluids of van der Waals character in section III obtaining exact results for the discretized critical points. Indeed, it was confirmed rather generally that the $\zeta$-dependence of $T_{\mbox{\scriptsize c}}(\zeta)$ and $\rho_{\mbox{\scriptsize c}}(\zeta)$ reflects closely the behavior of the second virial coefficient. Furthermore, the smoothness of the potential, $\varphi(r)$, or more precisely of the Mayer function, $f(r)=e^{-\varphi(r)/k_{\mbox{\tiny B}}T} -1 $, governs the improved rate of convergence of the critical parameters when $f(r)$ is continuous. Thus $d=1$ systems with a hard-core potential and jump in $f(r)$, generally exhibit only a slow, $1/\zeta$, convergence of $T_{\mbox{\scriptsize c}}(\zeta)$ and $\rho_{\mbox{\scriptsize c}}(\zeta)$ to their $\zeta\rightarrow\infty$ limits. However, if $\zeta$ is chosen properly --- in general {\em not} as an integer --- to match the second virial coefficient, $B_{2}^{\zeta}$, to its continuum limit, $B_{2}^{\infty}$, then the convergence becomes faster by a factor $1/\zeta$. Furthermore, the relative deviations of the critical parameters from the continuum $T_{\mbox{\scriptsize c}}$ and $\rho_{\mbox{\scriptsize c}}$ values become much smaller: see sections III.B-D. When $f(r)$ and its derivatives are continuous, the convergence rate is improved by further factors of $1/\zeta$: see sections III.D and E. 

The exact results for the one-dimensional models provide a useful guide to what may be expected in higher dimensions. However, understanding the behavior of more realistic two- and three-dimensional models requires further considerations that are presented in section IV and confirmed numerically via various simulation studies. From the results of section II, one learns that the second virial coefficients for models with hard-core potentials vary in an {\em intrinsically} erratic fashion as $\zeta$ increases. Thus even though the overall magnitude of the truncation error, $E_{0}(\zeta)$, decays as $1/\zeta^{(d+1)/2}$ (except for a logarithmic factor when $d=3$), its wildly erratic behavior severely hampers extrapolation of the critical parameters to their continuum limits for model fluids with hard-core potentials. To improve matters, a smoothed discretization parameter, $\zeta^{\dag}(\zeta)$, was introduced that serves to smooth the erratic critical data in a controlled way and thereby ease extrapolation to the continuum limit. A flexible procedure for {\em rescaling} $\rho$ and $T$ on the basis of $B_{0}^{\zeta}$ was also demonstrated successfully using simulations for the restricted primitive model electrolyte.

On the other hand, it was shown that it is feasible to select special values for $\zeta$ that make $E_{0}(\zeta)$ vanish: see Table 1. By adopting such optimal values one may hope to obtain critical point estimates lying sufficiently close to the continuum values that the need for extrapolation to $\zeta$$\,=\,$$\infty$ will be obviated in many practical cases.

\hspace{.2in}{\large \bf Acknowledgments}
\indent

We are indebted to Professor Pavel M.\ Bleher for directing us to important sources in the number theoretic literature and to him, Professor Freeman J.\ Dyson and Professor Joel L.\ Lebowitz for insightful and helpful comments. The interest and encouragement of Professor A.\ Z.\ Panagiotopoulos has been appreciated. Funding for our researches by the Department of Energy Office of Basic Energy Sciences (through Grant No.\ DE-FG0201ER15121 to A.\ Z.\ Panagiotopoulos, Princeton University) and by the National Science Foundation (through Grant No.\ CHE 03-01101 to M.E.F.) is gratefully acknowledged.

\appendix
\section{Relating Virial Coefficients to Critical Parameters}
\setcounter{equation}{0}
\renewcommand{\theequation}{A\arabic{equation}}
\indent

If eq \ref{eq1}, the rather general but approximate equation of state of van der Waals character, is accepted as a reliable guide, a direct proportionality between the rate of convergence of the discretized second virial coefficients, as defined in eq \ref{eq.4}, and the critical parameters, $T_{\mbox{\scriptsize c}}(\zeta)$, $\rho_{\mbox{\scriptsize c}}(\zeta)$, etc., must be expected as the discretization level, $\zeta$, approaches infinity.

However, this conclusion has a much broader basis. To see this, consider the Mayer fugacity expansion
 \begin{equation}
  p/k_{\mbox{\scriptsize B}}T = z + \sum_{l=2}^{\infty} b_{l}(T) z^{l},  \label{eq.A1}
 \end{equation}
or the virial expansion
 \begin{equation}
  p/k_{\mbox{\scriptsize B}}T = \rho + \sum_{l=2}^{\infty} B_{l}(T) \rho^{l}, \label{eq.A2}
 \end{equation}
and recall$^{25}$ that for a system with pair interaction potential $\varphi(\mbox{\boldmath $r$})$ the coefficients $b_l (T)$ and $B_l (T)$ can be defined explicitly as multiple integrals over products of Boltzmann factors $\exp[-\varphi(\mbox{\boldmath $r$}_{ij})/k_{\mbox{\scriptsize B}}T]$. Of course, these results hold in all dimensions $d$ and can be extended to multicomponent systems, many-body interactions, etc. In a discretized system all the cluster integrals must be replaced by sums as in eq \ref{eq.4}.

The series in eqs \ref{eq.A1} and \ref{eq.A2} and their discretized versions converge at small $z$ and $\rho$ for models of practical interest and in principle, by analytic continuation, then yield $T_{\mbox{\scriptsize c}}$ and $\rho_{\mbox{\scriptsize c}}$. For example, for systems displaying nonclassical critical behavior, which are of prime interest here, it suffices to examine the inverse compressibility, $1/K_{T}(T,\rho)$, which vanishes {\em uniquely} at $T=T_{\mbox{\scriptsize c}}$ and $\rho=\rho_{\mbox{\scriptsize c}}$. (Spinodal divergences of $K_T$ do not arise in realistic models owing to the essential singularities encountered at points of condensation$^{26}$ where the compressibility remains finite.)

As $\zeta$ increases the discretized second virial coefficient $B_{2}^{\zeta}(T)$ will approach its limit, say, with an exponent $\psi_B$. Unless accidental cancelations occur, one must, via eqs \ref{eq.A1} and \ref{eq.A2}, expect that the pressure and all thermodynamic quantities derived from $p(T,z)$ or $p(T;\rho)$ in the usual ways by differentiation, etc., can converge when $\zeta\rightarrow\infty$ {\em no more rapidly} than do the additive terms $~b_{2}^{\zeta}(T)z~$ and $~B_{2}^{\zeta}(T)\rho~$ [where, of course,$^{24}$ $B_{2}^{\zeta}=-b_{2}^{\zeta}\,\,$]. Conversely, unless the Boltzmann factors for any many-body potentials, which enter the expansions at higher order in $z$ and $\rho$, vary more sharply, e.g., with derivative discontinuities of lower order than in $\exp[-\varphi(\mbox{\boldmath $r$})/k_{\mbox{\scriptsize B}}T]$, the higher-order multiple discretized cluster integrals should {\em not introduce any slower convergence factors}. Indeed, further integration will generally have a smoothing effect leading to enhanced rates of convergence: see the dependence on dimensionality $d$ discussed in Appendix B. We thus conclude that, in general, the convergence exponents $\psi_T$ and $\psi_\rho$ match $\psi_B$ (or in very special cases --- see the discussion related to Table I --- may exceed $\psi_B$).

These arguments might seem to side-step the issue of nonclassical criticality raised in the Introduction. The crucial point, however, is that the second virial coefficient (discretized or not) closely reflects and, essentially, {\em embodies} the interaction potential $\varphi(r)$. But, as found explicitly even in highly nontrivial exactly solved models with nonclassical critical behavior --- in either one dimension$^{27}$ or in higher dimensions$^{28}$ --- the critical temperature $T_{\mbox{\scriptsize c}}$ varies smoothly and analytically with the potentials. Likewise for $\rho_{\mbox{\scriptsize c}}$ when this is not trivially fixed by some symmetry, etc. A well known example, the plane triangular Ising model$^{28}$ with three, positive, negative or vanishing exchange couplings, $J_1$, $J_2$ and $J_3$, illustrates this well and {\em also} indicates how the assertion may {\em fail} in special cases! Specifically, for certain particular values of the potentials the nature of the critical point itself may abruptly change. This happens in the triangular Ising model with $J_1, J_2, J_3 <0$ where an anomalous critical point arises {\em at} $T=0$ when $J_1 = J_2 = J_3$. In general, such points are {\em multicritical points}.$^{29}$

If a discretization process is performed on a system in the close neighborhood of any multicritical point, one must anticipate that the $\zeta$-dependence of the second virial coefficient, $B_{2}^{\zeta}(T)$, may be an unreliable guide to the behavior of the critical parameters. Indeed, near a bicritical point, for example, there are critical loci, say $T_{\mbox{\scriptsize c}}(g)$, which vary as $|g-g_{\mbox{\scriptsize b}}|^{\phi}$, where $g_{\mbox{\scriptsize b}}$ is the bicritical value of a potential parameter, $g$, while $\phi$ is a nonintegral (and nontrivial) crossover exponent.$^{30}$ Of course, in the situations addressed in this article, the implicit assumption --- that we may highlight here --- is that the critical point under consideration is, as typical in applications, one that resides in a manifold of critical points with uniform (and universal) critical behavior.

The insight into the nature of multicriticality reveals the fallacy in what might otherwise appear to be a cogent counter-example to our basic conclusion.$^{31}$ Suppose, recognizing explicitly the functional dependence on the interaction potential as embodied in the second virial coefficient, one regards the pressure as given by
 \begin{equation}
   p = {\cal P}\mbox{\boldmath $($}\rho,T;B_{2}^{\zeta}(T)\mbox{\boldmath $)$}.  \label{eq.A3}
 \end{equation}
Then for nonclassical criticality, the dependence on the first two arguments at $(\rho_{\mbox{\scriptsize c}},T_{\mbox{\scriptsize c}})$ must be nonanalytic; but one might suppose that the third argument also enters in some singular fashion. Then, from the vanishing of $1/K_T (T,\rho)$ or by eliminating $\rho$ between the standard conditions in eq 17, one might expect to derive a relation for $T_{\mbox{\scriptsize c}}(\zeta)$ of the form
 \begin{equation}
  {\cal K}[T_{\mbox{\scriptsize c}}(\zeta);\, B_{2}^{\zeta}\mbox{\boldmath $($} T_{\mbox{\scriptsize c}}(\zeta)\mbox{\boldmath $)$}] = 0.  \label{eq.A4}
 \end{equation}

Now if ${\cal K}(x;y)$ is analytic in both arguments at and through $x=x_{\mbox{\scriptsize c}}\equiv T_{\mbox{\scriptsize c}}^{\infty}$ and $y=y_{\mbox{\scriptsize c}}\equiv B_{2}^{\infty}(T_{\mbox{\scriptsize c}}^{\infty})$, our previous conclusion that the behavior of $[T_{\mbox{\scriptsize c}}(\zeta)-T_{\mbox{\scriptsize c}}^{\infty}]$ reflects $[B_{2}^{\zeta}(T_{\mbox{\scriptsize c}})-B_{2}^{\infty}(T_{\mbox{\scriptsize c}})]$ is confirmed. Conversely, if ${\cal K}(x;y)$ is singular in $y$ at $y_{\mbox{\scriptsize c}}$, perhaps varying as $(y-y_{\mbox{\scriptsize c}})^{\chi}$ with $\chi\neq 1$, the conclusion would generally be false: the convergence exponent $\psi_T$ for $T_{\mbox{\scriptsize c}}$ would rather be expected to depend on some combination of $\psi_B$ and $\chi$. However, the defect of this argument is that it presupposes, in essence, that the critical point in question at $(T_{\mbox{\scriptsize c}}^{\infty},\rho_{\mbox{\scriptsize c}}^{\infty})$ or $(x_{\mbox{\scriptsize c}},y_{\mbox{\scriptsize c}})$ is in reality a special critical point sensitive to the specific potential that yields $B_{2}(T_{\mbox{\scriptsize c}}^{\infty})=y_{\mbox{\scriptsize c}}$. As explained, this violates our basic presupposition that, in truth, must be required of any general scheme of analysis such as concerns us.

While we believe our arguments are convincing as regards the leading asymptotic dependence of $T_{\mbox{\scriptsize c}}(\zeta)$ and $\rho_{\mbox{\scriptsize c}}(\zeta)$ matching that of $B_{2}^{\zeta}(T)$, one may still ask whether the higher order corrections when $\zeta\rightarrow\infty$ might not depend in some way on the bulk critical exponents. It is not implausible that the answer is sensitive to the details as to how precisely one interprets the argument of $B_{2}^{\zeta}$: i.e., whether as supposed in eq \ref{eq.A4} or in some other reasonable way. A renormalization group approach and general scaling ansatz (such as employed in the articles by Kim and Fisher cited in ref 11) should be able to cast light on this issue. But, in the absence of some concrete and practical application, the point seems too academic to be worth pursuing at this juncture.

\section{Approximate Estimation of the Hard-core Truncation Error}
\setcounter{equation}{0}
\renewcommand{\theequation}{B\arabic{equation}}
\indent

Consider a $d$-dimensional sphere of radius $R=\zeta a_{0}$ centered at the origin of a discretized space with a lattice spacing $a_{0}$. Each lattice site may be regarded as the center of an elementary cell of volume $a_{0}^{d}$. The surface area of the sphere is $A_{R}=C_{d}R^{d-1}$ where $C_{d}d=v_{d}$ is the volume of a unit sphere. This surface then intersects the elementary lattice cells and divides them into two parts. The elements of the surface which are cut by individual lattice cells have a mean area, $A_{0}\simeq q_{d}a_{0}^{d-1}$ for $\zeta=R/a_{0}\gg 1$ where $q_{d}$ is a real number which we expect to remain of order unity for all $\zeta$. The mean number of such surface elements is thus given by
 \begin{equation}
   S_{R} \simeq A_{R}/A_{0} = (C_{d}/q_{d})\zeta^{d-1}.  \label{eq.B1}
 \end{equation}

For $d>1$ each surface element cuts a lattice cell at a different mean tangential angle and, depending on the location of the individual cell, its center will lie either just inside or just outside the sphere. When, overall, the surface elements are uncorrelated around the sphere, one may expect the probability of the center of a surface cell lying in each region to approach $\frac{1}{2}$ as $\zeta\rightarrow\infty$. Consequently, each more or less independent surface cell contributes a deviation or ``error'' of $+\frac{1}{2}$ or $-\frac{1}{2}$ to the total number of enclosed lattice points, $N_{d}(\zeta)$. The total mean square error should then be proportional to the number of the lattice cells cut by the surface of the sphere and so given by
  \begin{equation}
   \langle \Delta N_{d}^{2}\rangle = \langle n_{d}^{2}\rangle \approx S_{R}\delta_{d}^{2}, \label{eq.B2}
  \end{equation}
where $\Delta N_{d}(\zeta)= N_{d}(\zeta)-V_{d}(\zeta)=n_{d}(\zeta)$. The root-mean-square deviation $\delta_{d}$ should be of rough magnitude $\frac{1}{2}$ although {\em correlations} between nearby cells should increase the actual value. On using eq \ref{eq.B1}, we thus conclude that the root mean square error should vary as
 \begin{equation}
   n_{d}^{\mbox{\scriptsize rms}}(\zeta) = \sqrt{\langle n_{d}^{2}\rangle} \approx \delta_{d}\sqrt{C_{d}/q_{d}}\,\zeta^{(d-1)/2}.  \label{eq.B3}
 \end{equation}
Finally, this implies that the truncation error, $E_{0}(\zeta)=N_{d}(\zeta)/V_{d}(\zeta)-1$, should decay as $1/\zeta^{(d+1)/2}$.

As remarked, this result is (rather trivially) exact for $d=1$ while for $d=2$ it is confirmed by Bleher {\em et al.}$^{15}$ For $d=3$, however, the exact results$^{14}$ imply that an additional logarithmic factor, $\ln\zeta$, should appear in eq \ref{eq.B2}. Furthermore, the argument fails for $d>3$ since $n_{d}(\zeta)$ grows like $\zeta^{d-2}$ implying$^{12,14}$ $E_{0}(\zeta)\sim 1/\zeta^{2}$.

To understand why there should be a borderline or crossover dimension we may, first, note that if, in $d$ dimensions we had considered {\em oriented hypercubic hard cores} of sides $2R$ (rather than spheres) a very different result would pertain. Indeed, a $d$-hypercube can be decomposed into $2d$ rectangular hypercones of opening angle $\theta=\pi/4$ about each lattice axis. The face (or ``base'') of each pyramidal hypercone would be of area $A_{d}=(2R)^{d-1}=(2a_{0})^{d-1}\zeta^{d-1}$ and on increasing $R$ by one lattice spacing $a_{0}$, the number of enclosed lattice sites would increase discontinuously in a completely correlated manner by, in total,
 \begin{equation}
   \Delta N_{d}(\zeta) \approx 2d (A_{d}/a_{0}^{d-1}) = 2^{d}d\,\zeta^{d-1}. \label{eq.B4}
 \end{equation}
On dividing by the hypercube volume $V_{d}=2^{d}R^{d}$, the rms truncation error is seen to be $E_{0}(\zeta) \propto d/\zeta$ and so decays like $1/\zeta$ for all $d$.

This transparent result suggests that for sufficiently large $d$ the growing surface of a hypersphere may, from the perspectives of the $2d$ outgoing lattice axes, resemble a somewhat flat base of a cone of opening angle, say, $\theta\leq \pi/4$. Let us, to be specific, focus on the positive $x$ axis and consider the lattice layer at or closest to the (normal) distance $X=R\cos\theta$ from the sphere center and of transverse dimension $R_{\perp}=R\sin\theta$. Then if $d\geq 3$ and $X$ increases by a lattice spacing $a_{0}$, a {\em fully correlated} set of sites contained in a ``ring'' or torus of thickness $\Delta X=a_{0}$, width $\Delta Y\simeq a_{0}\cot\theta$ (provided $\theta^{2}\gtrsim 1/\zeta$) and ``rectangular circumference'' of magnitude $(2R_{\perp})^{d-2}$ is added to $N_{d}(\zeta)$. Thus fluctuations in $N_{d}(\zeta)$ of magnitude at least
 \begin{equation}
   \Delta N_{d}^{\theta}(\zeta) \propto a_{0}^{2}R^{d-2}(\sin\theta)^{d-3}\cos\theta/a_{0}^{d}\propto \zeta^{d-2}  \label{eq.B5}
 \end{equation}
must be anticipated. But, clearly, this implies that $E_{d}(\zeta)$ cannot decrease faster than $1/\zeta^{2}$ for $d\geq 3$, just as the rigorous results establish! Incidently, formally maximizing $\Delta N_{d}^{\theta}$ on $\theta$ yields $\:\theta=\pi/4\:$ for $d\geq 5$, $\:\theta=\sin^{-1}(3^{-1/2})\simeq 35.3^{\circ}\:$ for $d=4$, and $\:\theta=\theta_{\mbox{\scriptsize min}}\gtrsim 1/\zeta^{1/2}\:$ for $d=3$.

\pagebreak
\hspace{-.2in}{\large \bf References and Notes}
\begin{itemize}
 \item[(1)] Fisher, M.\ E.\ {\em J.\ Stat.\ Phys.} {\bf 1995}, {\em 75}, 1.
 \item[(2)] Panagiotopoulos,  A.\ Z.; Kumar, S.\ K.\ {\em Phys.\ Rev.\ Lett.} {\bf 1999}, {\em 83}, 2981.
 \item[(3)] Panagiotopoulos, A.\ Z.\ {\em J.\ Chem.\ Phys.} {\bf 2000}, {\em 112}, 7132. It should be noted that the gain in speed afforded by the fine-discretization technique can be sensitive to programming details, machine architecture, compilation parameters, etc.
 \item[(4)] Panagiotopoulos, A.\ Z.\ {\em J.\ Chem.\ Phys.} {\bf 2002}, {\em 116}, 3007.
 \item[(5)] Moghaddam, S.; Panagiotopoulos, A.\ Z.\ {\em J.\ Chem.\ Phys.} {\bf 2003}, {\em 118}, 7556.
 \item[(6)] Luijten, E.; Fisher, M.\ E.; Panagiotopoulos, A.\ Z.\ {\em Phys.\ Rev.\ Lett.} {\bf 2002}, {\em 88}, 185701.
 \item[(7)] Cheong, M.\ D.\ W.; Panagiotopoulos, A.\ Z.\ {\em J.\ Chem.\ Phys.} {\bf 2003}, {\em 119}, 8526.
 \item[(8)] Kim, Y.\ C.; Fisher, M.\ E.\ {\em Phys.\ Rev.\ Lett.} {\bf 2004}, {\em 92}, 185703. Note that the definition of the truncation errors, $E_{i}(\zeta)$, introduced in this reference differs slightly from eq \ref{eq.5} here in that the ratio $B_{i}^{\infty}/B_{i}^{\zeta}$ is employed there in place of its reciprocal. Of course, no difference in the rates of convergence results.
 \item[(9)] One must, however, note that at high densities and in any crystalline phase, the presence of an underlying periodic lattice may, and typically will, induce spurious commensurate-incommensurate phase transitions not characteristic of the system in the full continuum limit. Furthermore, for low values of $\zeta \lesssim 3$ such effects are much stronger and may lead to qualitatively different phase diagrams even at moderate densities: see ref 2. However, for the investigation of fluid phases at densities up to $\frac{1}{2}\rho_{\mbox{\scriptsize max}}$ or somewhat larger, such features do not arise in the simple molecular models for $\zeta\gtrsim 4$; thus they will not be a concern for us here.
 \item[(10)] It may be remarked here, and is touched on briefly below in section IV, that periodic boundary conditions, which one normally wishes to impose, must respect the periodicity of the fine lattice of spacing $a_{0}$. Thus, if $L$ is a side of a rectangular simulation box parallel to a lattice axis, $L/a_{0}$ must be an integer. Alternatively, in terms of the oft-used notation, $L^{\ast}\equiv L/a$, one must require that $L^{\ast}\zeta$ is integral.
 \item[(11)] Of course, it is important to realize that the critical parameters $T_{\mbox{\scriptsize c}}(\zeta)$ and $\rho_{\mbox{\scriptsize c}}(\zeta)$ are not, themselves, directly open to evaluation via simulations. Rather, it is crucial to simulate for a range of system sizes $L$ and to extrapolate in an unbiased and systematic fashion in order to gain reliable, but inevitably somewhat uncertain, estimates of $T_{\mbox{\scriptsize c}}(\zeta)$ and $\rho_{\mbox{\scriptsize c}}(\zeta)$ for $L\rightarrow\infty$. In such extrapolations, allowance for the nonclassical values of the exponents $\beta$, $\gamma$, $\nu$, etc., is crucial. Some recent theoretical and numerical studies bearing directly on this issue are: Orkoulas, G.; Fisher, M.\ E.; Panagiotopoulos, A.\ Z.\ {\em Phys.\ Rev.\ E} {\bf 2001}, {\em 63}, 051507; Luijten, E.; Fisher, M.\ E.; Panagiotopoulos, A.\ Z.\ {\em Phys.\ Rev.\ Lett.} {\bf 2002}, {\em 88}, 185701; Kim, Y.\ C.; Fisher, M.\ E.; Luijten, E.\ {\em Phys.\ Rev.\ Lett.} {\bf 2003}, {\em 91}, 065701; Kim, Y.\ C.; Fisher, M.\ E.\ {\em Phys.\ Rev.\ E} {\bf 2003}, {\em 68}, 041506; {\em J.\ Phys.\ Chem.\ B} {\bf 2004}, {\em 108}, 6750 and ref 8.
 \item[(12)] Walfisz, A.\ {\em Gitterpunkte in mehrdimensionalen Kugeln}; PWN-Polish Scientific Publishers: Warsaw, 1957.
 \item[(13)] While the proofs we cite are mathematically rigorous, our reports here do not explicitly acknowledge the suitably weighted smoothing of the errors $E_{i}(\zeta)$, over a range of $\zeta$ values, that is essential in formulating precise analytical results.
 \item[(14)] Bleher, P.\ M.; Dyson, F.\ J.\ {\em Acta Arith.} {\bf 1994}, {\em 68}, 383; erratum: {\bf 1995}, {\em 73}, 199.
 \item[(15)] Bleher, P.\ M.; Dyson, F.\ J.; Lebowitz, J.\ L.\ {\em Phys.\ Rev.\ Lett.} {\bf 1993}, {\em 71}, 3047.
 \item[(16)] Bleher, P.\ M.\ {\em Emerging Applications of Number Theory}, The IMA Volumes in Mathematics and its Applications, Vol.\ 109; Springer-Verlag: New York, 1999, p.\ 1-38.
 \item[(17)] Fisher, M.\ E.; Widom, B.\ {\em J.\ Chem.\ Phys.} {\bf 1969}, {\em 50}, 3756.
 \item[(18)] See, e.g., Fisher, M.\ E.\ in unpublished work reported at the Second Eastern Theoretical Physics Conference, Chapel Hill, N.C., 25 October 1963 and at Institut des Hautes \'{E}tudes Scientifiques, Bures-sur-Yvette, Paris, 26 April 1965.
 \item[(19)] Kac, M.; Uhlenbeck, G.\ E.; Hemmer, P.\ {\em J.\ Math.\ Phys.} {\bf 1963}, {\em 4}, 216, 299; Lebowitz, J.\ L.; Penrose, O.\ {\em ibid}, {\bf 1963}, {\em 4}, 248.
 \item[(20)] Lennard-Jones, J.\ E.\ {\em Proc.\ R.\ Soc.\ London, Ser.\ A} {\bf 1924}, {\em 106}, 441; {\em ibid}, {\bf 1924}, {\em 106} 463.
 \item[(21)] Buckingham, R.\ A.\ {\em Proc.\ R.\ Soc.\ London, Ser.\ A} {\bf 1938}, {\em 168}, 264.
 \item[(22)] The Buckingham exponential-6 potential$^{21}$ (which replaces the repulsive part of the Lennard-Jones potential by an exponential function) has a maximum at $r=r_{\mbox{\scriptsize max}}\simeq 0.2285\sigma$, and then diverges sharply to $-\infty$ when $r\rightarrow 0$ owing to the $-1/r^{6}$ term. Because of this anomalous behavior, simulations (and our calculations here) have employed a modified form of the original E-6 potential obtained by introducing a hard-core, $\varphi(r)=+\infty$, for $r<r_{\mbox{\scriptsize max}}$. 
 \item[(23)] Errington, J.\ R.; Panagiotopoulos, A.\ Z.\ {\em J.\ Chem.\ Phys.} {\bf 1998}, {\em 109},  1093.
 \item[(24)] See, e.g., McQuarrie, D.\ A.\ {\em Statistical Mechanics}; Harper Collins: New York, 1976, section 15-2, and Friedman, H.\ L.\ {\em Ionic Solution Theory based on cluster expansion methods}; Interscience Publishers: New York, 1962.
 \item[(25)] Mayer, J.\ E.; Mayer, M.\ G.\ {\em Statistical Mechanics}; John Wiley: New York, 1940, chapter 13, and, e.g., McQuarrie, D.\ A.\ of ref 24, chapter 12.
 \item[(26)] See Fisher, M.\ E.\ {\em Physica A} {\bf 1990}, {\em 163}, 15; in ``{\em Proc.\ Gibbs Symposium, Yale University, May 1989},'' Eds.\ Caldi, D.\ G.; Mostow, G.\ D.; Amer.\ Math.\ Soc.: Rhode Island, 1990; and references therein.
 \item[(27)] Fisher, M.\ E.; Felderhof, B.\ U.\ {\em Ann.\ Phys.\ (N.Y.)} {\bf 1970}, {\em 58}, 176, 217; Felderhof, B.\ U.; Fisher, M.\ E.\ {\em ibid}, {\bf 1970}, {\em 58}, 268; Felderhof, B.\ U.\ {\em ibid}, {\bf 1970}, {\em 58}, 281.
 \item[(28)] See, e.g., Baxter, R.\ J.\ {\em Exactly Solved Models in Statistical Mechanics}; Academic Press: London, 1982.
 \item[(29)] Fisher, M.\ E.\ In {\em Proceedings of the 1974 AIP Conference No.\ 24, Magnetism and Magnetic Materials}; American Institute of Physics: New York, 1975, p.\ 273; Aharony, A.\ in {\em Critical Phenomena}, Lecture Notes in Physics No.\ 186, Ed.\ Hahne, F.\ J.\ W.; Springer: Berlin, 1983, p.\ 209.
 \item[(30)] For critical loci near bicritical points, see Fisher, M.\ E.; Nelson, D.\ R.\ {\em Phys.\ Rev.\ Lett.} {\bf 1974}, {\em 32}, 1350; Fisher, M.\ E.\ {\em Phys.\ Rev.\ Lett.} {\bf 1975}, {\em 34}, 1634.
 \item[(31)] Proferred by an anonymous referee.
\end{itemize}

\pagebreak
\begin{table}

\caption{Selected matching values of $\zeta\equiv a/a_{0}$ in the range $(5,25)$ (recorded to ten decimal places) for $d=2$ and $3$ dimensions. Using these values in simulations of fluids with hard-cores of diameter $a$, should enhance the degree of approximation of the continuum model by the discretized version.}
\vspace{0.3in}
\begin{tabular}{r|r}
   $d=2~~~~~$ & $d=3~~~~~$ \\ \hline 
 $5.89031$$\;$$21813~$ &$~  5.97830$$\;$$60184$ \\
 $6.79374$$\;$$22306~$ &$~  6.86800$$\;$$19972$ \\
 $7.83797$$\;$$21888~$ &$~  7.94786$$\;$$02312$ \\
 $8.75857$$\;$$76568~$ &$~  8.89198$$\;$$85870$ \\
 $9.65737$$\;$$00691~$ &$~  9.96022$$\;$$01803$ \\
$10.89631$$\;$$07310~$ &$~ 10.91631$$\;$$51776$ \\
$11.79412$$\;$$65154~$ &$~ 11.93598$$\;$$08041$ \\
$12.82833$$\;$$62583~$ &$~ 12.92528$$\;$$08130$ \\
$14.89503$$\;$$24158~$ &$~ 14.93315$$\;$$20912$ \\
$16.78404$$\;$$74639~$ &$~ 16.93406$$\;$$53012$ \\
$18.72055$$\;$$51383~$ &$~ 18.93232$$\;$$51547$ \\
$20.81393$$\;$$17548~$ &$~ 20.94656$$\;$$23713$ \\
$22.91054$$\;$$34749~$ &$~ 22.95780$$\;$$06533$ \\
$24.85637$$\;$$72316~$ &$~ 24.95754$$\;$$85659$ \\
\end{tabular}
\end{table}
\pagebreak
\begin{figure}
\begin{center}
 Figure Captions
\end{center}
\caption{Schematic illustration of the fine-lattice discretization process for $\zeta=1$, $2$, $3$, and $5$ for a $(d\,$$=$$\,2)$-dimensional system where $a$ is the hard-core diameter while $a_{0}$ is the lattice spacing. The dashed circle of radius $a=\zeta a_{0}$ encloses the sites, shown shaded, excluded by hard-core repulsion from occupation by neighboring particles. In $d$$\,=\,$$3$ dimensions these numbers are are $N_{3}(\zeta)=1,27,93,251$ and $485$ for $\zeta=1$ to $5$.}

\caption{Relative hard-core truncation errors, $E_{0}(\zeta)$, scaled by $\zeta^{(d+1)/2}$ for (a) a one-dimensional discrete system and (b) a $d=2$ simple square lattice.}

\caption{The distribution, ${\cal P}(s)$, of the fluctuating variable, $s\equiv [N_{d}(\zeta)-V_{d}(\zeta)]/\zeta\sqrt{\ln\zeta}$, for $d=3$ calculated numerically from $\zeta=5$ up to $1000$ using an increment $\Delta\zeta = 10^{-4}$ and a bin size $\Delta s = 0.1$. Even for $\zeta\ge 10$ the distribution is very close to the limiting Gaussian distribution that is labeled $\zeta=\infty$.}

\caption{Plots of (a) $\Delta\rho_{\mbox{\scriptsize c}}(n)/\rho_{\mbox{\scriptsize c}}(\infty)$ and (b) $\Delta T_{\mbox{\scriptsize c}}(n)/T_{\mbox{\scriptsize c}}(\infty)$ vs $1/n$ for one-dimensional square-well models of range $c=\lambda a$ with $\lambda =1\frac{1}{2}$, $2$, $3$ and $5$ where the discretization parameter $\zeta$ takes integer values.}

\caption{Plots of (a) $\Delta\rho_{\mbox{\scriptsize c}}(\zeta)/\rho_{\mbox{\scriptsize c}}(\infty)$ and (b) $\Delta T_{\mbox{\scriptsize c}}(\zeta)/T_{\mbox{\scriptsize c}}(\infty)$ vs $1/\zeta^{2}$ for one-dimensional square-well models with $\lambda = 1\frac{2}{3}$, $2\frac{1}{3}$, $3$ and $5$ where the discretization parameter runs through the nonintegral values $\zeta=n-\frac{1}{2}$ for $\lambda = 3$ and $5$ but only $\zeta = 3(n-\frac{1}{2})$ for $\lambda=1\frac{2}{3}$ and $2\frac{1}{3}$.}

\caption{The Boltzmann factor, $e^{-\beta\varphi(r)}$, vs distance for (a) one-dimensional logarithmic hard-core models (solid lines) with $\lambda = 2$ and $w=0$, $\frac{1}{2}$, $\frac{3}{4}$ and $2$ at a temperature set by $\beta\epsilon = 1$; the dotted lines are for the hard-core square-well models with $\lambda= 1\frac{1}{2}$, $2$ and $3$. (b) Boltzmann factor for a cubic model at $\beta\epsilon = 1$ with the parameters given in the text.}
\end{figure}
\pagebreak
\newpage
\begin{figure}
\caption{Plots of (a) $\Delta\rho_{\mbox{\scriptsize c}}(n)/\rho_{\mbox{\scriptsize c}}(\infty)$ and (b) $\Delta T_{\mbox{\scriptsize c}}(n)/T_{\mbox{\scriptsize c}}(\infty)$ vs $1/n$ for one-dimensional logarithmic hard-core models with $\lambda = 2$ and $w=0$, $0.75$, and $2$ when the discretization parameter $\zeta$ takes integer values. The insets show the behavior of the critical temperature and density when $w=0$ as functions of $1/n^{2}$: see Figure 6.}

\caption{Plots of (a) $\Delta\rho_{\mbox{\scriptsize c}}(\zeta)/\rho_{\mbox{\scriptsize c}}(\infty)$ and (b) $\Delta T_{\mbox{\scriptsize c}}(\zeta)/T_{\mbox{\scriptsize c}}(\infty)$ vs $1/\zeta^{2}$ for the $d=1$ logarithmic repulsive-core models with $\lambda = 2$ and $w=0.5$, $0.75$ and $2$ for the choice $\zeta=n-\frac{1}{2}$ with $n=5$, $6$, $\cdots\;$.}

\caption{Plots of (a) $\Delta\rho_{\mbox{\scriptsize c}}(n)/\rho_{\mbox{\scriptsize c}}(\infty)$ and (b) $\Delta T_{\mbox{\scriptsize c}}(n)/T_{\mbox{\scriptsize c}}(\infty)$ vs $1/n^{4}$ for a one-dimensional ``cubic'' model (see text) where $\zeta$ takes successive integer values demonstrating rapid convergence for smooth potentials in $d=1$ dimensions.}

\caption{The Boltzmann factor, $e^{-\beta\varphi(r)}$, vs distance for the Lennard-Jones $(12,6)$ potential (solid curve) and for the modified Buckingham exponential-6 potential (dashed curve) evaluated at the estimated critical temperatures, $k_{\mbox{\scriptsize B}}T_{\mbox{\scriptsize c}}/\epsilon\simeq 1.299$ and $1.243$, respectively.$^{3,23}$}

\caption{The reduced second virial coefficients, $B_{2}^{\ast}(T_{\mbox{\scriptsize c}};\zeta)$ vs $1/\zeta^{6}$ for (a) the $(12,6)$ Lennard-Jones (LJ) and (b) modified Buckingham exponential-6 (E-6) potentials in $d$$\,=\,$$3$ dimensions. The values have been calculated at the estimated continuum critical temperatures (see Figure 10) for $5\leq \zeta \leq 25$. In simulations using the E-6 potential estimates of $T_{\mbox{\scriptsize c}}(\zeta)$ have been observed$^{5}$ to approach the limit as $1/\zeta^{6\pm2}$. The insets show the same data at a larger scale vs $1/\zeta^{2}$ (see also the text).}

\caption{Plots of (a) $T_{\mbox{\scriptsize c}}^{\ast}(\zeta)$ and $\rho_{\mbox{\scriptsize c}}^{\ast}(\zeta)$ for the RPM electrolyte vs $1/\zeta^{\dagger\, 2}$ where $\zeta^{\dagger}(\zeta)$ is the smoothed discretization parameter calculated from the hard-core second virial coefficient: see eq \ref{eq.E0} and the following text. The symbols from the left correspond to the choices $c^{\dagger}=35E_{0}(5)$, $25E_{0}(5)$, and $15E_{0}(5)$. The dashed lines are to guide extrapolation to $\zeta=\infty$; the data come from ref 8.}

\end{figure}
\pagebreak
\newpage

\begin{figure}
\centerline{\epsfig{figure=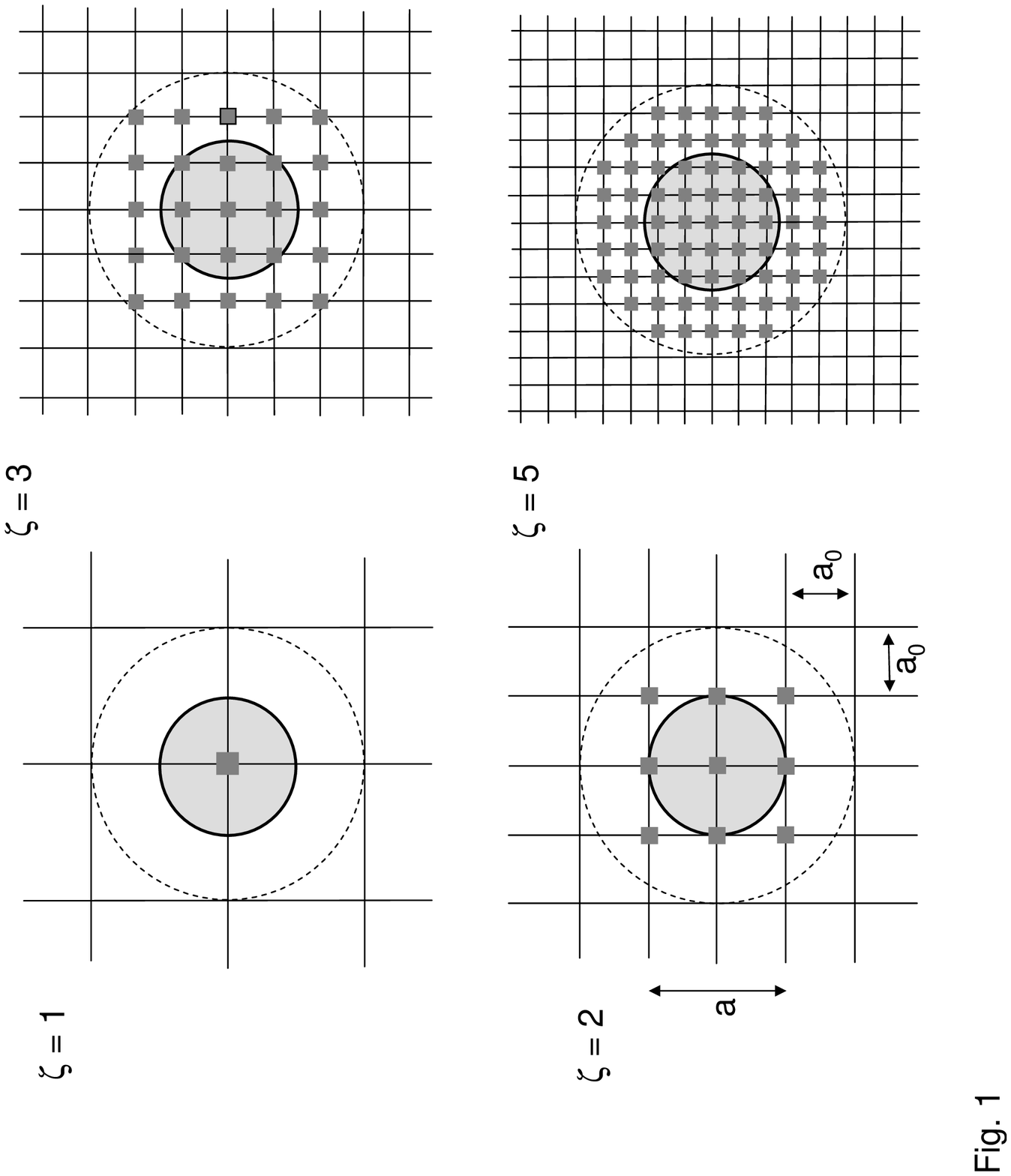,width=8.0in,angle=0}}
\end{figure}
\begin{figure}
\centerline{\epsfig{figure=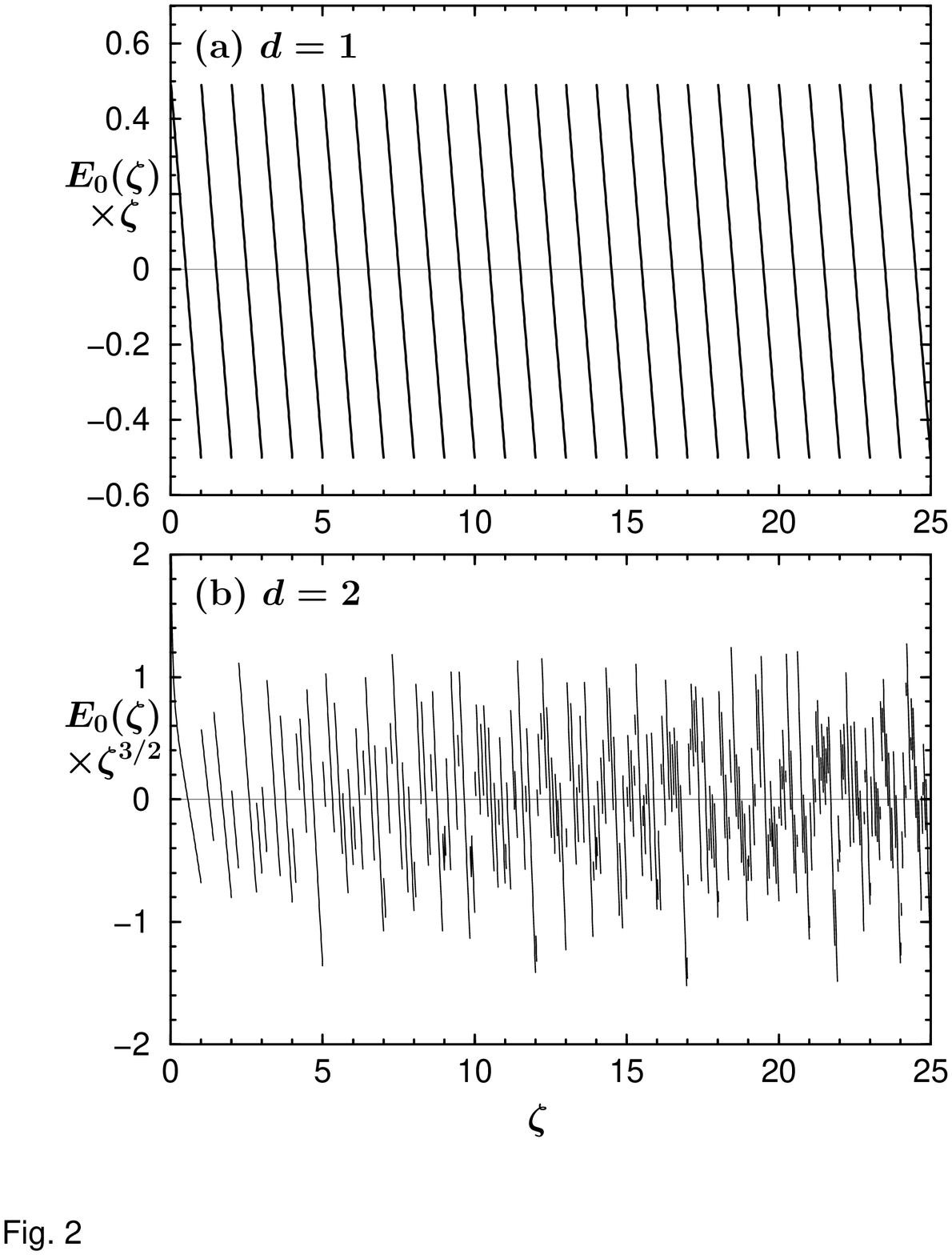,width=7.3in,angle=0}}
\end{figure}
\begin{figure}
\centerline{\epsfig{figure=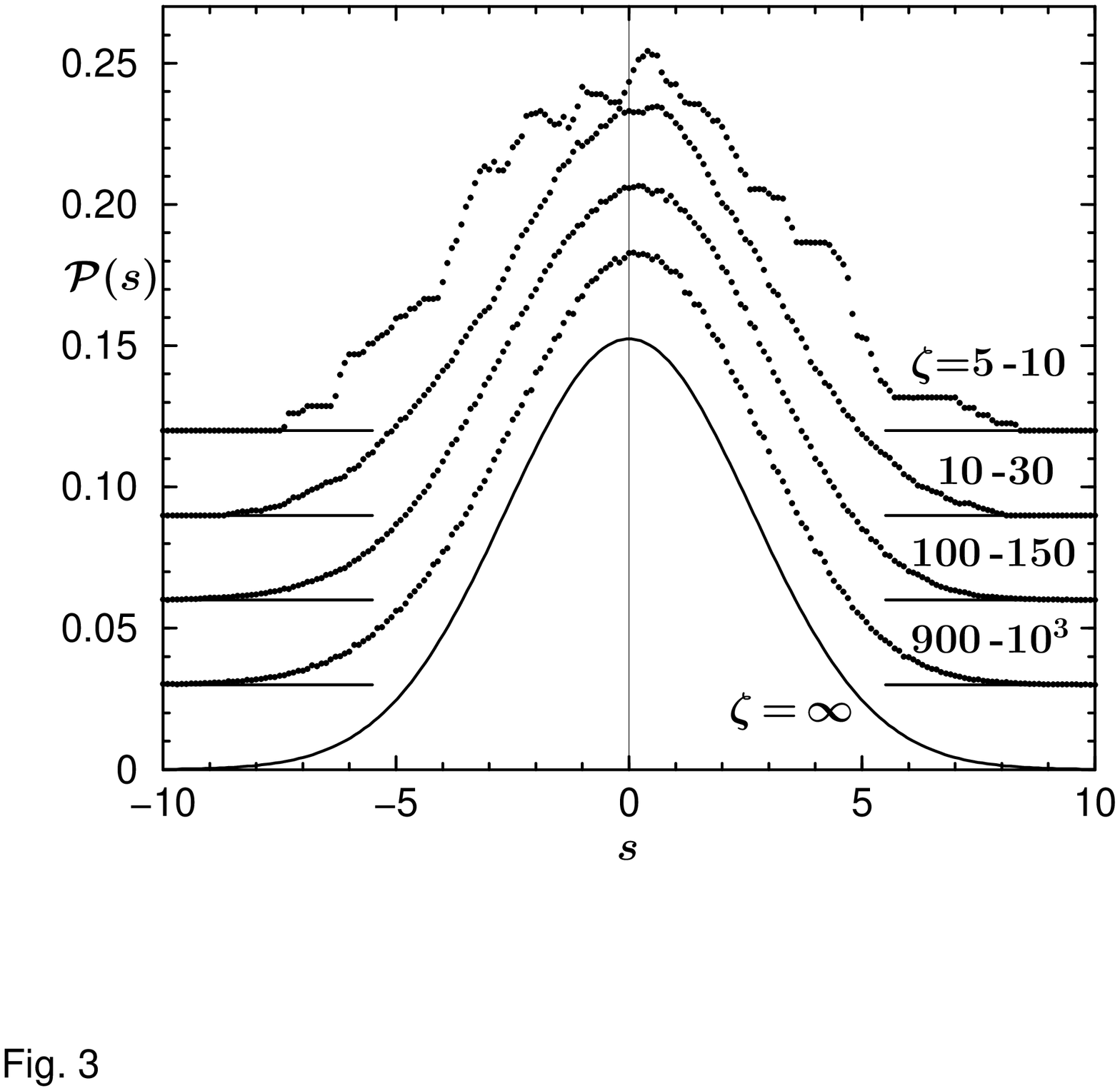,width=7.3in,angle=0}}
\end{figure}
\begin{figure}
\centerline{\epsfig{figure=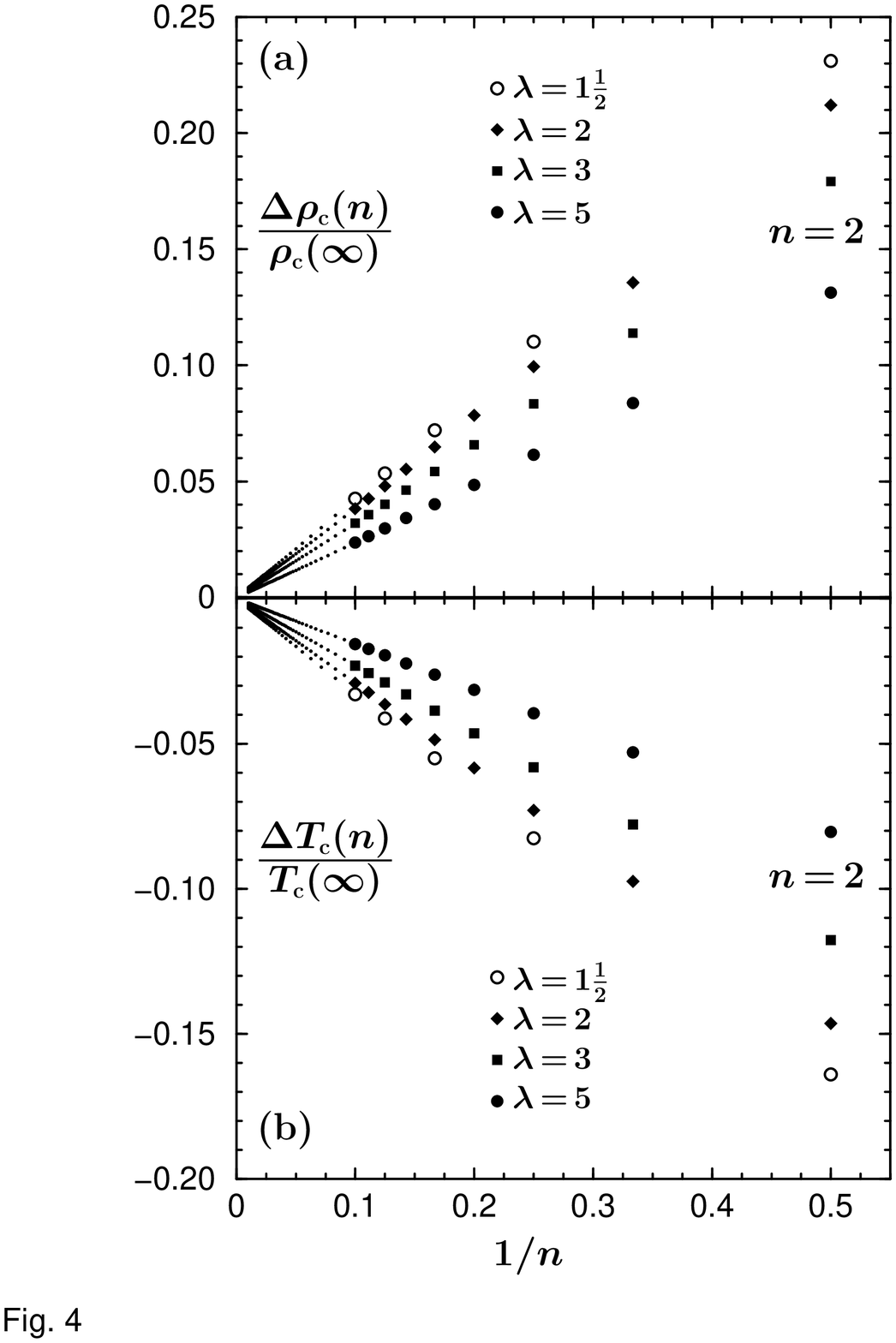,width=7.3in,angle=0}}
\end{figure}
\begin{figure}
\centerline{\epsfig{figure=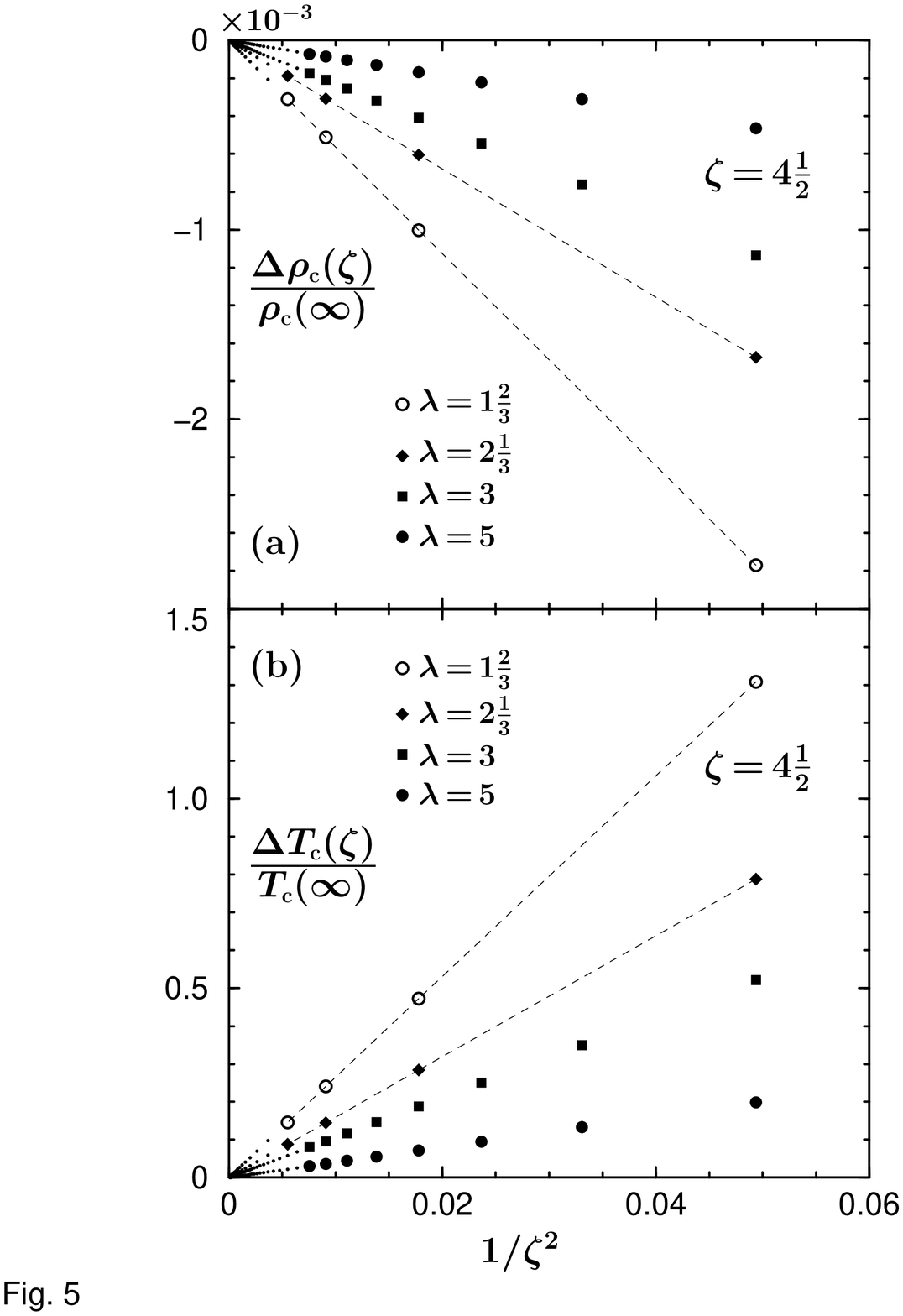,width=7.3in,angle=0}}
\end{figure}
\begin{figure}
\centerline{\epsfig{figure=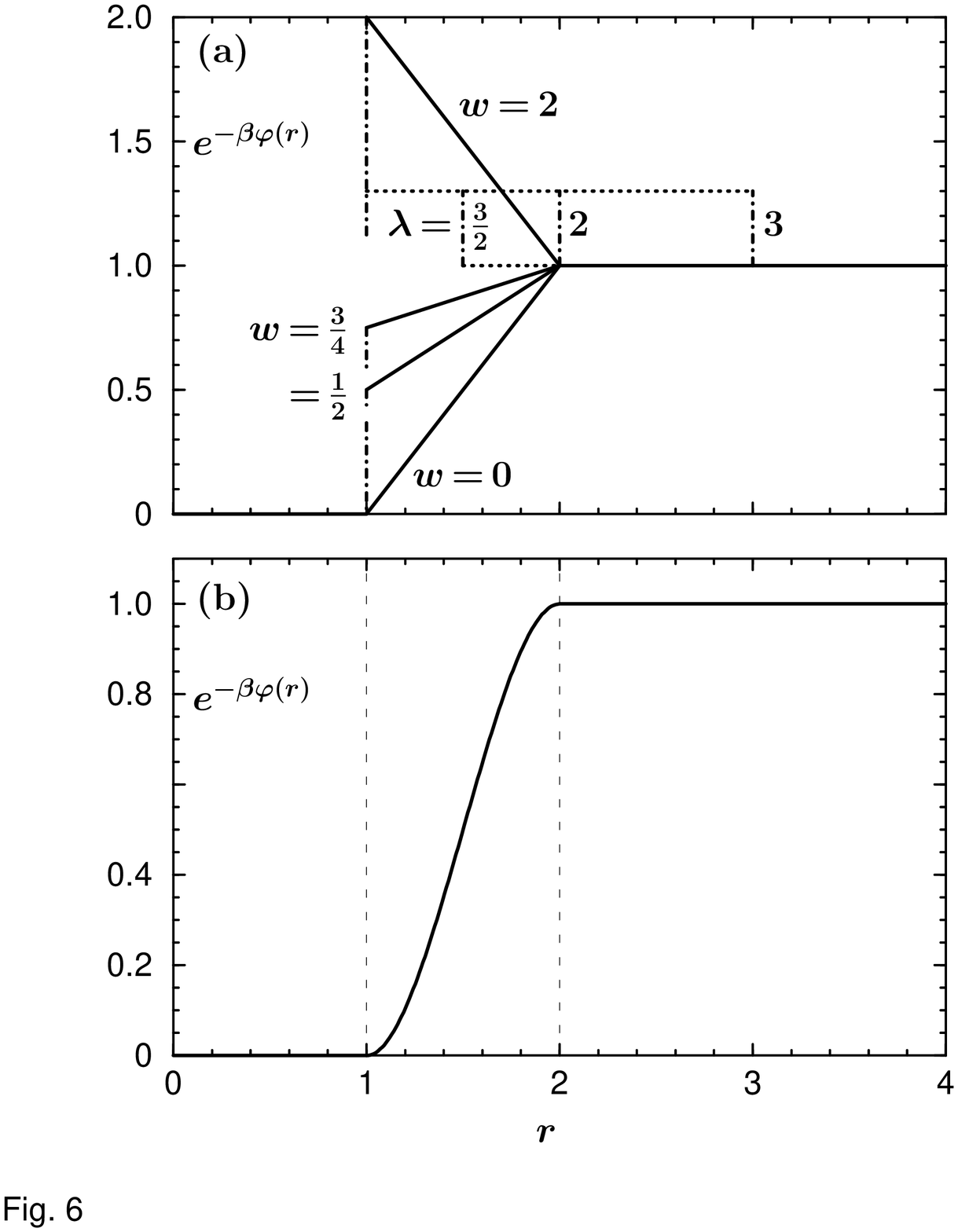,width=7.3in,angle=0}}
\end{figure}
\begin{figure}
\centerline{\epsfig{figure=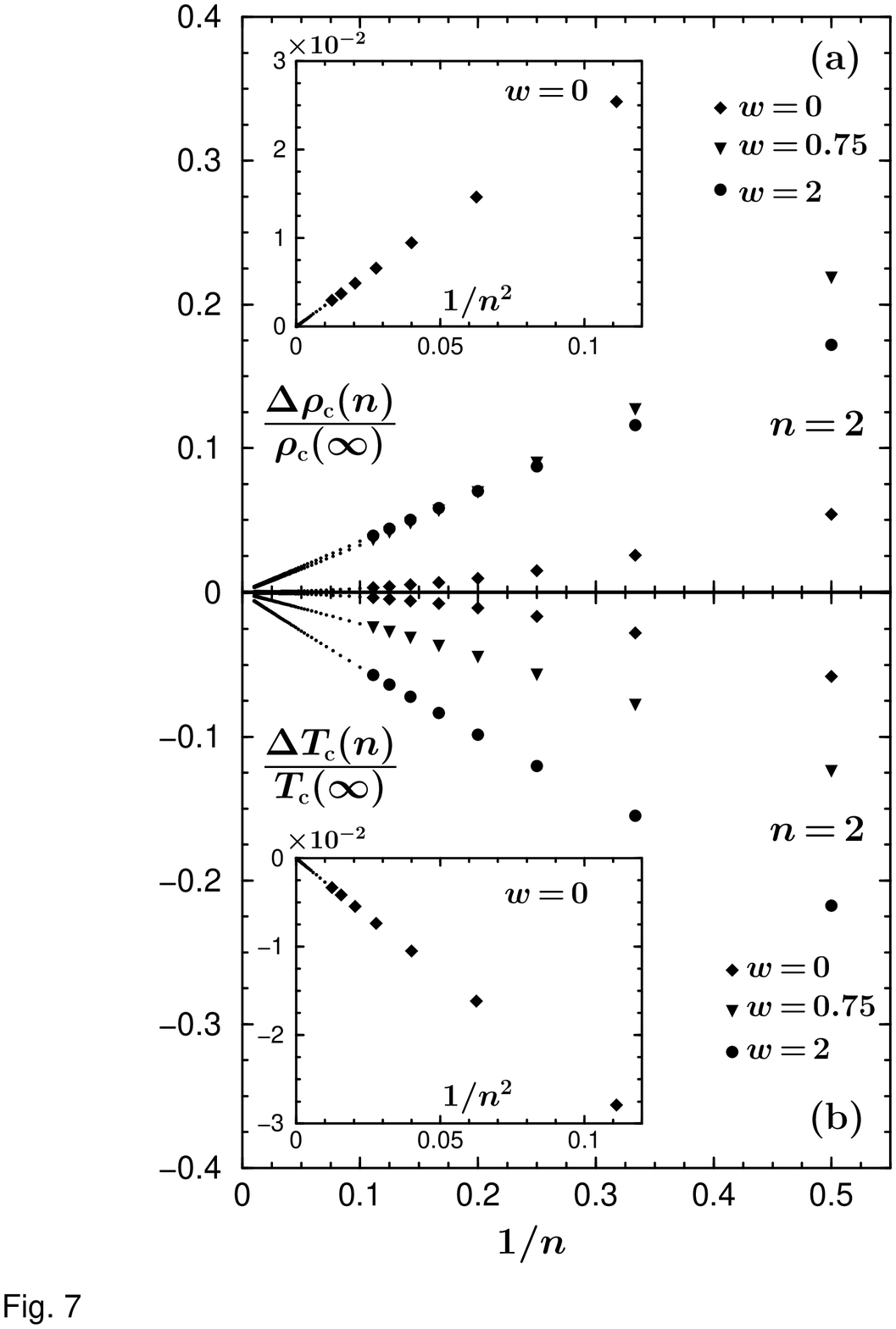,width=7.3in,angle=0}}
\end{figure}
\begin{figure}
\centerline{\epsfig{figure=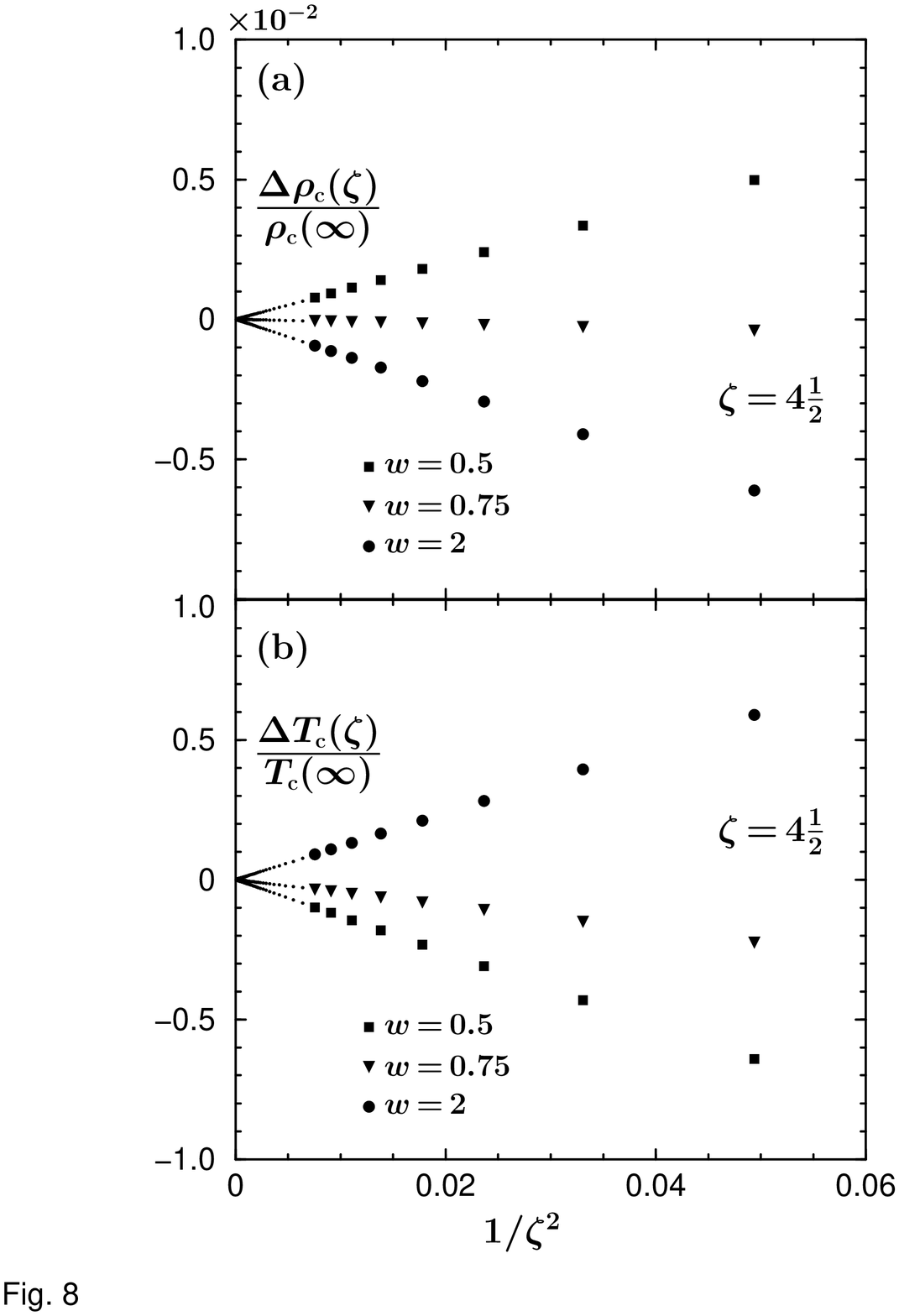,width=7.3in,angle=0}}
\end{figure}
\begin{figure}
\centerline{\epsfig{figure=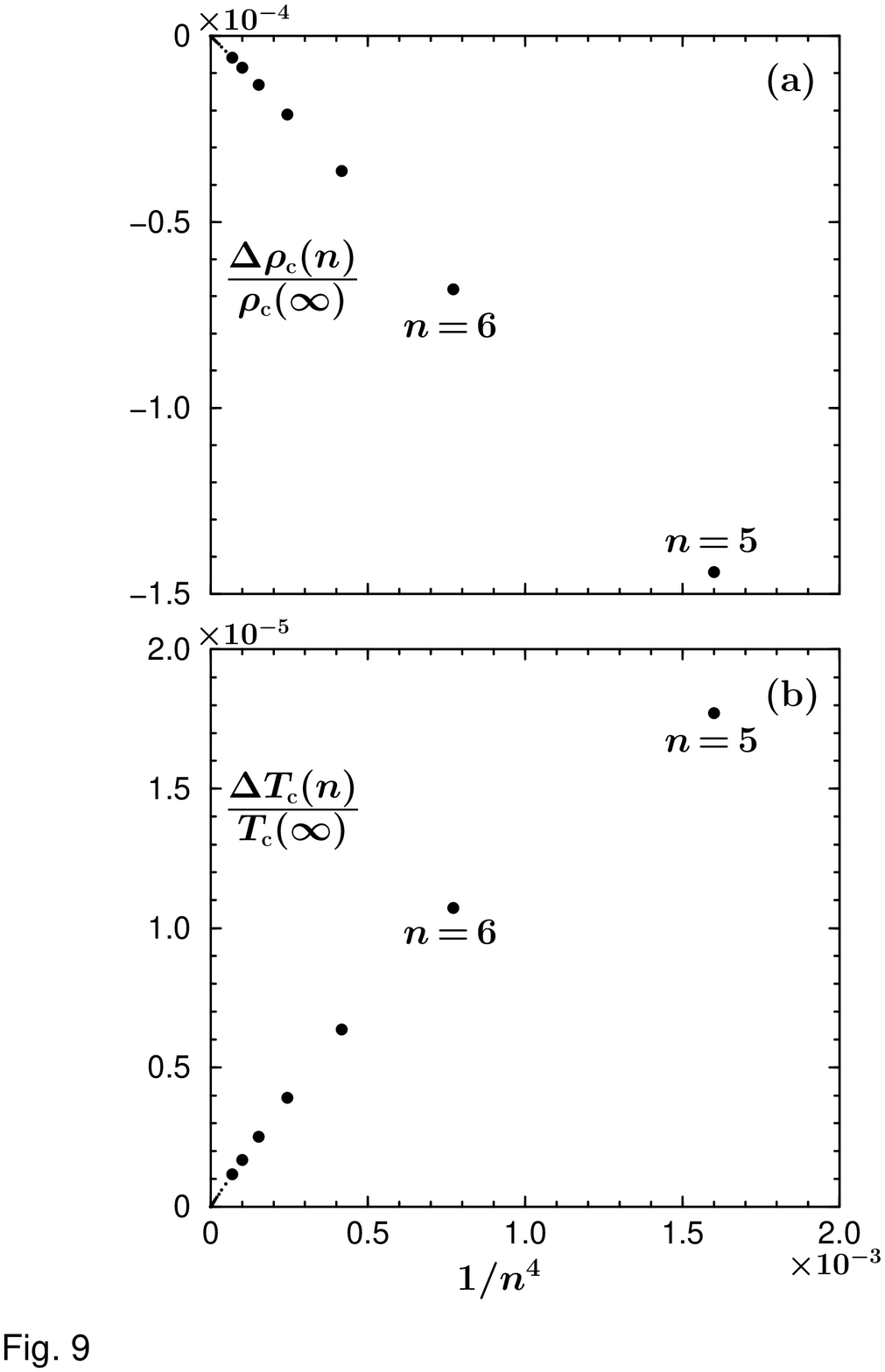,width=7.3in,angle=0}}
\end{figure}
\begin{figure}
\centerline{\epsfig{figure=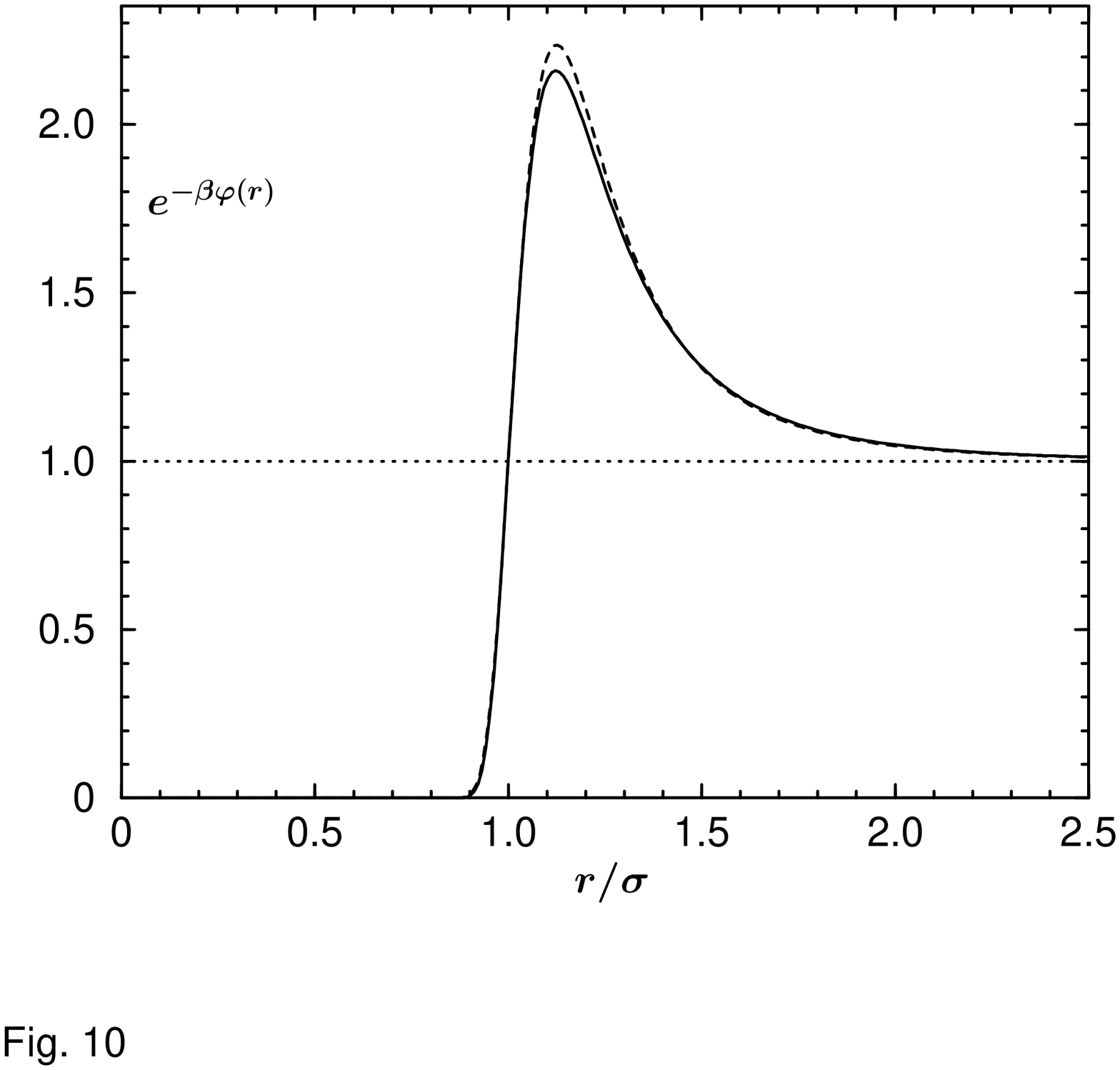,width=7.3in,angle=0}}
\end{figure}
\begin{figure}
\centerline{\epsfig{figure=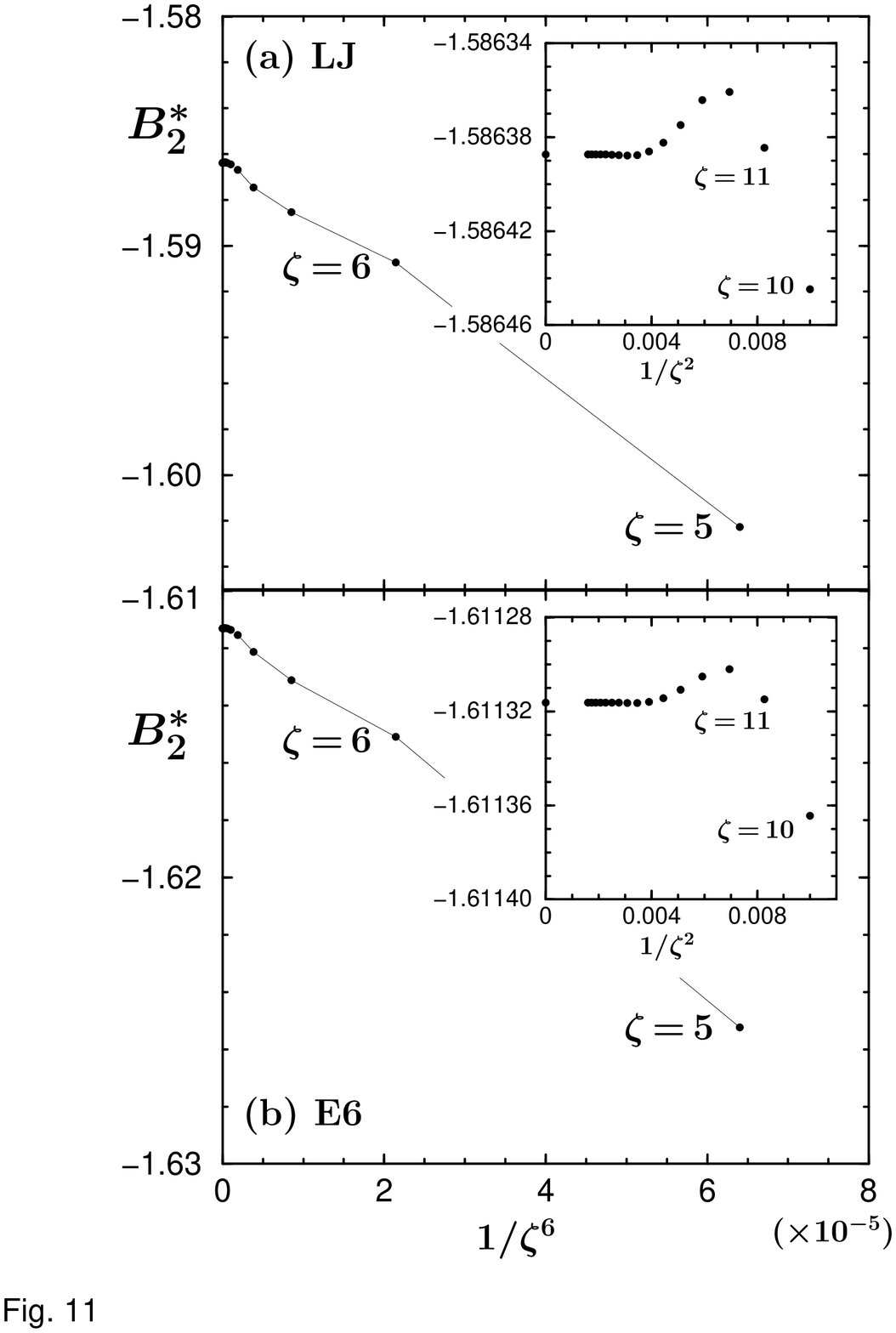,width=7.3in,angle=0}}
\end{figure}
\begin{figure}
\centerline{\epsfig{figure=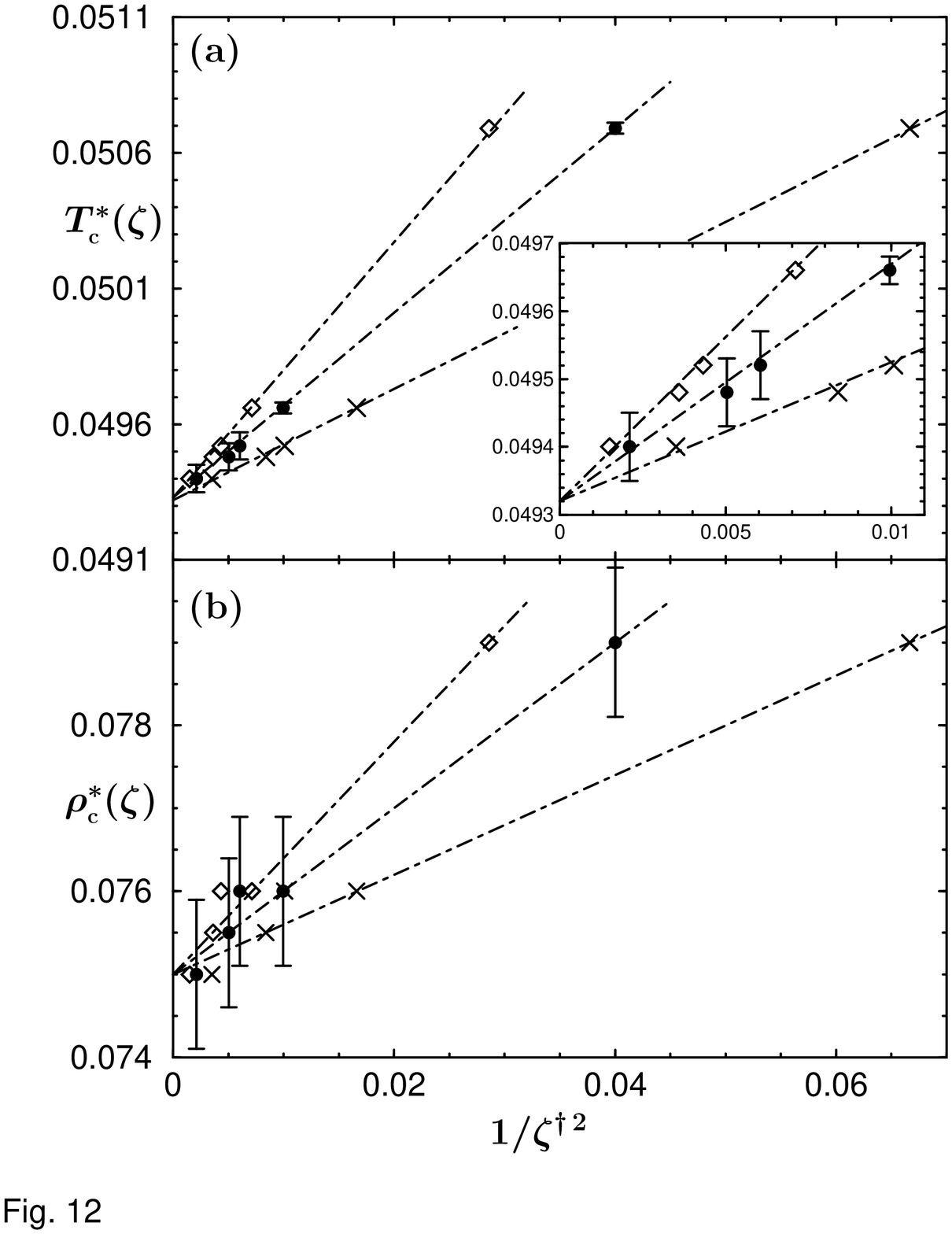,width=7.3in,angle=0}}

\end{figure}

\end{document}